\documentclass[12pt]{spieman}  % 12pt font required by SPIE;
\usepackage{amsmath,amsfonts,amssymb}
\usepackage[square,sort,comma,numbers]{natbib}
\usepackage{graphicx}
\usepackage{setspace}
\usepackage{tocloft}
\usepackage{color}
\usepackage[inkscapelatex=false]{svg}
\usepackage{caption}
\usepackage{subcaption}
\usepackage{graphicx}
\usepackage{float}
\usepackage{flafter}
\usepackage[T1]{fontenc}

\title{Domain Transfer Through Image-to-Image Translation for Uncertainty-Aware Prostate Cancer Classification}

\author[a,b,*]{Meng Zhou}
\author[b]{Amoon Jamzad}
\author[c]{Jason Izard}
\author[c]{Alexandre Menard}
\author[c]{Robert Siemens}
\author[b,**]{Parvin Mousavi}
\affil[a]{University of Toronto, Department of Computer Science, 40 St George St., Toronto, ON, Canada, M5S 2E4}
\affil[b]{Queen’s University, Medical Informatics Laboratory, 557 Goodwin Hall, Kingston, ON, Canada, K7L 2N8}
\affil[c]{Kingston Health Sciences Research Centre, 76 Stuart Street, Kingston, ON, Canada, K7L 2V7}

\cftpagenumbersoff{figure}
\cftpagenumbersoff{table} 
\begin{document} 
\maketitle

\begin{abstract}
\textbf{Purpose: }Prostate Cancer (PCa) is a prevalent disease among men, and multi-parametric MRIs offer a non-invasive method for its detection. While MRI-based deep learning solutions have shown promise in supporting PCa diagnosis, acquiring sufficient training data, particularly in local clinics remains challenging. One potential solution 
is to take advantage of publicly available datasets to pre-train deep models and fine-tune them on the local data, but multi-source MRIs can pose challenges due to cross-domain distribution differences. These limitations hinder the adoption of explainable and reliable deep-learning solutions in local clinics for PCa diagnosis. In this work, we present a novel approach for unpaired image-to-image translation of prostate multi-parametric MRIs and an uncertainty-aware training approach for classifying clinically significant PCa, to be applied in data-constrained settings such as local and small clinics. \textbf{Method: }Our approach involves a novel pipeline for translating unpaired 3.0T multi-parametric prostate MRIs to 1.5T, thereby augmenting the available training data. Additionally, we introduce an evidential deep learning approach to estimate model uncertainty and employ dataset filtering techniques during training. Furthermore, we propose a simple, yet efficient \textit{Evidential Focal Loss}, combining focal loss with evidential uncertainty, to train our model effectively. \textbf{Results: }Our experiments demonstrate that the proposed method significantly improves the Area Under ROC Curve (AUC) by over 20\% compared to the previous work (98.4\% vs. 76.2\%). \textbf{Conclusions: }Our proposed framework effectively translates and aligns public data with local data to increase the number of training data for deep models. Additionally, our proposed uncertainty estimation method enhances PCa detection performance. Providing prediction uncertainty to radiologists may aid in prioritizing uncertain cases, and expediting the diagnostic process effectively. Our code is available at~\url{https://github.com/med-i-lab/DT_UE_PCa}.
\end{abstract}

% Include a list of up to six keywords after the abstract
\keywords{Deep Learning, Image Translation, Uncertainty Estimation, Prostate Cancer}

% Include email contact information for corresponding author
{\noindent \footnotesize\textbf{*}The majority of this work was completed while M.Zhou was at the Medical Informatics Laboratory, Queen's University.}

{\noindent \footnotesize\textbf{**}Parvin Mousavi, corresponding author,   \linkable{mousavi@queensu.ca}

\begin{spacing}{2}   % use double spacing for rest of manuscript

\section{Introduction}
\label{sect:intro}  % \label{} allows reference to this section
Prostate Cancer (PCa) is a prevalent form of cancer among men \cite{reda2018deep}, and the clinically significant PCa is defined by the Gleason score $>$ 6 or the histopathology ISUP grade $\geq$ 2 \cite{smith2004identification, arif2020clinically}. The current PCa diagnosis procedure involves a combination of the prostate-specific antigen test and the histopathology analysis of the Transrectal Ultrasound-guided biopsy (TRUS) taken from 10-12 regions on the prostate gland \citep{2019, grebenisan2021spatial}. However, the histopathology analysis on TRUS can miss up to 20\% of clinically significant PCa due to the limited number of biopsy samples \citep{grebenisan2021spatial,reda2018deep}.

% Multi-parametric Magnetic Resonance Imaging (mp-MRI) has emerged as an effective alternative to TRUS for the early detection of PCa. mp-MRI uses a combination of anatomical and functional sequences of MRI that can further highlight the differences between normal and abnormal (cancer) cells. 
% 3.0T MRI generally has higher image quality and spatial resolution \citep{ladd2018pros} than 1.5T MRI, but the latter is widely used in local, small clinical centers due to its lower price \citep{cost_1997}.

Multi-parametric Magnetic Resonance Imaging (mp-MRI) has emerged as an effective alternative to TRUS for the early detection of PCa. mp-MRI uses a combination of anatomical and functional sequences of MRI that can further highlight the differences between normal and abnormal (cancer) cells. The evaluation and reporting guideline of prostate mp-MRI was first introduced in the Prostate Imaging Reporting and Data System (PI-RADS) \citep{barentsz2012esur, weinreb2016pi, barentsz2016synopsis}. The guideline provides a comprehensive scoring schema for suspicious prostate lesions and mp-MRI sequences. An extensive Prostate MRI imaging study (PROMIS) \citep{bosaily2015promis} reported that targeted biopsy using mp-MRI has higher sensitivity and negative-predictive value (NPV) but lower specificity compared to TRUS biopsy \citep{bosaily2015promis, stabile2020multiparametric, ahmed2017diagnostic}. The study also showed that 27\% of the patients did not need to undergo biopsy, had mp-MRI been used for screening. Although PROMIS provides strong practical implications for mp-MRI in PCa diagnosis, the low specificity indicates that mp-MRI can be plausibly improved by advanced analyses.

In recent years, deep learning methods have emerged as a powerful tool for image classification tasks, and have provided promising performance in detecting and segmenting PCa on multi-parametric Prostate MRIs \cite{saha2021end, le2017automated, yoo2019prostate, iqbal2021prostate,  pellicer2022deep}. A more recent grand challenge, ProstateX \cite{armato2018prostatex}, has further shown the ability of deep learning approaches in detecting clinically significant PCa on 3.0T mp-MRI data. Several groups have developed Convolutional Neural Network (CNN)-based models that achieve high performance for PCa classification \cite{litjens2014computer, liu2017prostate, mehrtash2017classification, armato2018prostatex, grebenisan2020towards, grebenisan2021spatial}. These methods have a great potential for clinical translations by highlighting abnormal lesions for radiologists during the PCa diagnostic process. 

While deep learning has shown promising results in detecting PCa on mp-MRI, there are several substantial challenges in training and deploying deep models in clinics. Training deep models typically requires a large amount of data, which is not always available for local clinics with limited patient throughput. The alternative is to use pre-trained models or publicly available labeled data for training. However, the effectiveness of these models is significantly impacted by the differences in distribution between the public models/data and the images from local centers \cite{grebenisan2021spatial}. The primary contributing factor to this disparity is the field strength of MRI scanners.

While 3.0T scanner generally has higher image quality and spatial resolution \cite{ladd2018pros} and most of the public datasets for prostate cancer are predominantly acquired with 3.0T scanners \cite{adams2022dataset,adams2022prostate158,hulsen2019overview}, 1.5T machines are considered standard of care and widely used in local clinical centers due to their cost-effectiveness \cite{cost_1997}. Despite the preference for 3.0T MRI scanners, PI-RADS guidelines state that both 1.5T and 3.0T can provide adequate and reliable diagnostic examinations \cite{PIRADS} and certain research indicates no statistically significant difference in clinical impact between 1.5T and 3.0T MRIs for diagnosis of prostate cancer \cite{ryznarova2018,virarkar2022,woernle2023}. Considering that over 80\% of MRI systems in Canada operate at a 1.5T while exhibiting a higher rate of exams per hour than their 3.0T counterparts \cite{CADTH24}, 1.5T MRIs continue to be the standard in clinical care. 

Therefore, to bridge the distribution gap between the public data and clinical application, a necessary step is to translate 3.0T MRI data to 1.5T. This translation process facilitates aligning the data distributions and allows for the training of classification models using both the translated data and local data. It is worth noting that there are ongoing efforts in the literature to address related challenges in federated learning \cite{li2020multi, adnan2022federated}. However, this aspect is not the focus of this study.

Besides, classical deep models are primarily designed for predicting labels during the inference of test data, irrespective of whether the test image is within or outside the training set distribution. These models lack the capability to discern data samples belonging to unrelated distributions \citep{sensoy2018evidential} or to quantify their confidence in predictions. These limitations make models hard to interpret, raising concerns about the reliability of such models. Hence, reusing and deploying models for local PCa detection is challenging. Addressing the above limitations and drawbacks is crucial when integrating deep learning models into real clinical routines. Thus, two main questions arise in this context:
%For example, prostate MRI typically with high magnetic strength of 3.0T \citep{ullrich2017magnetic} is preferred because it produces high-resolution images and provides detailed information. In contrast, the low magnetic strength MRI (1.5T) may result in fuzzy boundaries \citep{ladd2018pros} and not be able to offer detailed information.
\begin{enumerate}
    \item For small local clinical centers, can they take advantage of the extensive high-resolution 3.0T public MRI data to enhance the classification performance on their limited low-resolution local 1.5T MRI data?
    %For local small hospitals, how could they obtain a reliable AI model to detect PCa with limited data?
    \item When deploying models in clinical centers, can we provide supplementary information regarding the confidence of the model's predictions, beyond the final result, to enhance the reliability and explainability of the models?
    % When deploying models in clinical centers, could we offer additional information regarding the confidence of the model's predictions, in addition to the final result, to enhance the reliability of the models?
    %For radiologists, how could they know whether the AI model is confident in its prediction or not?
\end{enumerate}

In this work, we aim to answer the two important questions outlined above. To this end, we propose a novel 2-stage learning framework for clinically significant PCa classification using multi-parametric, multi-center MRI data. Our framework not only aims to enhance classification performance but also provides an estimate of predictive confidence alongside the corresponding predicted label, introducing a valuable dimension to the model's interpretability.
% that can simultaneously provide an estimate of the predictive confidence and the corresponding predicted label to improve classification performance.
In the first stage, we introduce a data preprocessing pipeline that translates prostate mp-MRI data from 3.0T to 1.5T via a Generative Adversarial Network (GAN) approach to increase the number of training samples. This step addresses the challenge of limited data in local clinics with low patient throughput (refer to Section \ref{sec:dt}). In the second stage, we propose an uncertainty-aware PCa classification approach. Specifically, we explore various model architectures to enhance classification performance, experiment with different strategies for combining multi-parametric MRI data, and leverage the \textit{co-teaching} framework \citep{han2018co} to mitigate potential issues related to noisy labels (refer to Section \ref{model_archis}). During the training phase, we incorporate dataset filtering using \textit{evidential uncertainty estimation} \citep{sensoy2018evidential} to eliminate training data samples with high prediction uncertainty, thereby enhancing the robustness of our models. Additionally, we extend the work of \cite{sensoy2018evidential} by introducing a novel \textit{Evidential Focal Loss} to optimize our classification models during training (see Section \ref{evi_focal}). The experimental results affirm the effectiveness of our proposed framework, demonstrating a significant improvement in classification performance compared to previous work. Our contributions not only advance the field of PCa detection but also underscore the potential of uncertainty-aware approaches in enhancing the reliability and interpretability of deep learning models in clinical settings.

\textbf{Contributions:} In summary, our work makes three main contributions:

\begin{enumerate}
  \item We develop a GAN-based framework to translate unpaired prostate mp-MRIs from 3.0T to 1.5T, which we termed as domain transfer. This framework would align different data distributions and increase the number of training data for deep classification models. This is also the first attempt to translate prostate mp-MRIs in an unpaired manner. 
  % \item We utilize the prediction uncertainty in the model outputs based on the Theory of Evidence \citep{Yager2008}, so that our model is able to say "I do not know" for highly uncertain data samples. We further filtering out highly uncertain data to improve the model robustness. We propose a novel loss function termed \textit{Evidential Focal Loss} that combines the original Focal Loss \citep{lin2017focal} and the evidential uncertainty \citep{sensoy2018evidential} for the PCa classification task.
  % \item We incorporate the Theory of Evidence \citep{Yager2008} into our model, enabling it to identify and filter out highly uncertain training data and making the model more robust. We propose a novel loss function termed \textit{Evidential Focal Loss}. To the best of our knowledge, it is the first time that the original Focal Loss \citep{lin2017focal} is combined with the evidential uncertainty \citep{sensoy2018evidential} for binary PCa classification.
  \item We propose a novel loss function termed \textit{Evidential Focal Loss} that can jointly compute the uncertainty for training samples and optimize the classification model. To the best of our knowledge, it is the first time that the original Focal Loss \citep{lin2017focal} is combined with the evidential uncertainty \citep{sensoy2018evidential} for binary PCa classification.
  
  \item By filtering the training samples based on their uncertainty value, our results outperform the state-of-the-art and improve the interpretability of model predictions. By providing confidence estimates for the predictions, radiologists can make informed decisions during the PCa diagnostic process and effectively expedite the process.
  % for radiologists during the PCa diagnostic process via uncertainty estimation.
\end{enumerate}

\section{Related Work}
\subsection{Domain Adaptation} \label{relw:da}

Machine learning algorithms usually perform well when training and test data share the same distribution and feature space. However, in real-world applications, the distribution of test data often shifts, leading to biased or inaccurate predictions. In addition, it is time-consuming or infeasible to acquire new training data and fully repeat training steps. Domain Adaptation (DA) is an approach that addresses this issue by mitigating the dataset bias or domain shift problem caused by different distributions. There has been a lot of work on this topic in the past few years, which can be grouped into the following three general tasks~\citep{cui2020heuristic}: (1) unsupervised DA tasks \citep{ganin2015unsupervised, ganin2016domain, long2016unsupervised, saito2018maximum, long2014transfer} focus on addressing the domain shift problem without requiring labeled target domain data; (2) semi-supervised DA tasks\citep{yao2015semi, saito2019semi, li2021learning} aim to explore the partially labeled target domain data to further enhance the performance of domain adaptation algorithms; and (3) multi-source DA tasks \citep{hoffman2012discovering, xu2018deep, peng2019moment} deal with scenarios where multiple source domains are available for adaptation. %DA algorithms have been widely applied to extract domain-invariant features \citep{zhao2019learning, li2021learning} for knowledge transfer between source and target domains. Methods used to tackle the above DA tasks are categorized into three general areas \citep{wang2018deep}%
DA methods are often used to extract domain-invariant features for transferring knowledge between source and target domains. These methods incorporate various learning objectives with deep neural networks \cite{wang2018deep} for distribution matching:
% DA algorithms are often used to extract domain-invariant features for knowledge transfer between source and target domains. These algorithms fall into three general categories:
(1). \textbf{Discrepancy Measurement-based} methods aim to align feature distributions between two domains by fine-tuning deep models, e.g., using statistic criterion like Maximum Mean Discrepancy \citep{long2015learning, yan2017mind, kumagai2019unsupervised}, and class criterion \citep{tzeng2015simultaneous, hinton2015distilling, motiian2017unified}. Some of these methods often require large labeled target domain data to diminish the domain shift problem, which is sometimes infeasible to get such medical data in the real-life scenario. (2). \textbf{Adversarial-based} methods aim to confuse domain discriminators from Generative Adversarial Networks (GANs) to enhance the invariant feature extraction \citep{ganin2016domain, bousmalis2017unsupervised, hong2018conditional}. One common scenario involves utilizing noise vectors, either with or without source images, to generate realistic target images while preserving the source features. However, training GANs are hard and sometimes results in generator degradation, e.g., mode collapse \citep{karras2020training}. (3). \textbf{Reconstruction-based} methods, in addition to the general GANs approach from the above category, aim to reconstruct source-like images as an auxiliary task to preserve domain invariant features through an adversarial reconstruction paradigm \citep{hoffman2018cycada, zhu2017unpaired}. These methods usually have superior performance over the conventional GANs approach because they have an explicit reconstruction task to supervise the entire pipeline and make the training process more stable.
%encoder-decoder, {zhuang2015supervised, bousmalis2016domain}

%Among all reconstruction-based methods, CycleGAN \citep{zhu2017unpaired} is one of the state-of-the-art models for unpaired image-to-image translation.
CycleGAN \citep{zhu2017unpaired} is one of the state-of-the-art unsupervised adversarial reconstruction-based methods that is widely used for unpaired image-to-image translation. Its cycle consistency loss ensures the pixel-level similarity between two images through a reconstruction task, i.e., the source image $s$ is translated to the target domain $\hat{s}$ and then translated back $\tilde{s}$, where it should be identical to the original image ($s = \tilde{s}$). However, a drawback of the cycle consistency loss lies in its stringent constraint on pixel-level similarity, which will degrade the performance of GANs in some certain tasks \citep{zhao2020unpaired}. To address this limitation, the adversarial consistency loss GAN (ACL-GAN) \cite{zhao2020unpaired} is proposed to replace the pixel-level similarity with the distance between distributions. This modification allows ACL-GAN to retain essential features from source images while overcoming the drawbacks associated with the strict cycle consistency constraint. Therefore, we adapt the ACL-GAN model and build our framework based on it.

% Recently, \cite{zhao2020unpaired} purposed the Adversarial Consistency Loss, ACL-GAN, an improved version of CycleGAN \citep{zhu2017unpaired} for the same task as CycleGAN. They state that the classical cycle-consistency loss \citep{zhu2017unpaired} is a harsh constraint on the pixel-level, which will degrade the performance of GANs. Accordingly, they aim to minimize the distance between distributions instead of minimizing the pixel differences. In this way, the ACL-GAN model will ensure the model could still retain important features from the source images and overcomes the disadvantage of the cycle-consistency constraint. 
%i.e., instead of forcing $s = \tilde{s}$, we let the distribution of $\tilde{s}$ to be similar to the distribution of $s$.

%\subsubsection{Domain Adaptation in Medical Imaging}

In medical imaging, domain shift problems usually fall into two variations: subject-related variation (age, gender, etc.), and acquisition-related variation (MRI vendor, field strength, imaging protocol, etc.) \citep{kouw2017mr}. To solve such problem, one intuitive approach is to fine-tune a model that is pre-trained on the source domain with the new data from the target domain. \cite{Khan2019} propose to use the pre-trained VGG model on the ImageNet dataset \citep{deng2009imagenet} to learn robust high-level features of natural images, and then fine-tune it on the labeled MR images for the Alzheimer’s Disease (AD) classification task to achieve state-of-the-art performance. %They further select the most informative training samples by using the image entropy. Finally% They evaluate their method on the ADNI dataset \citep{JackJr2008} and achieve the state-of-the-art performance. 
Similarly, \cite{Ghafoorian2017} study the impact of the fine-tuning techniques on the brain lesion segmentation task, demonstrating that fine-tuning with only a small number of target domain training samples can outperform models trained from scratch. % They use a CNN model pre-trained on legacy brain MR images as the base model.
% Experiments show that fine-tuning with only a small number of target domain training samples can outperform the similar model that trained from scratch.
Another approach is to use domain adaptation as an intermediate step to reduce variance in image acquisition parameters from both domains and then use it for downstream tasks. Researchers have attempted to address the problem of acquisition variation in MRI data for several years. \cite{kouw2017mr} propose a feature-level representation learning method to either extract acquisition-invariant features or remove acquisition-variant features from paired 1.5T and 3.0T brain MRIs. The learned features are then used for a downstream classification task. %They first utilize the Siamese network with paired 1.5T and 3.0T MRI data to classify whether two patches from the same region are similar or not, and then use the output to train a classifier.
However, obtaining paired 1.5T and 3.0T MRI data in real-life scenarios is impractical. Another way to align acquisition-invariant features is to synthesize images from different types of acquisition parameters using GAN-based adversarial reconstruction methods. GANs have been applied to perform cross-modality image translation between different medical images or generate synthetic images from random noise. The objective of such translation tasks is to retain the underlying structure while changing the appearance of the image \citep{armanious2020medgan}. Researchers have attempted to estimate images in the target modality from the source modality, such as MRI-CT translation \citep{hiasa2018cross, nie2017medical, oulbacha2020mri, armanious2019unsupervised} and X-ray to CT translation \citep{ying2019x2ct, ge2022x}. Other areas that have been explored include intra-modality translation, such as T1/T2-FLAIR translation \citep{hu2021multi, uzunova2020memory} and pure data augmentation by generating synthetic images from random noise vectors \citep{radford2015unsupervised, frid2018gan, huang2021enhanced, kwon2019generation}. However, most of the works do not consider the real clinical practicality, for example, T1/T2-FLAIR MRI translation may require paired training data, which is not feasible in real clinical settings. Generating synthetic images from noise does not take advantage of the publicly available data and ignores \textit{a-priori} information. The current limitations provide great potential for unpaired image translation for medical images, which we employ in this work. 
%3.0T-7.0T MRI translation \citep{nie2018medical}

\subsection{Deep Learning for PCa Classification}
%{\color{red}More ref (about prostateX), less detail of each}
%DC maps, maximum b-value from DWI and $K^{trans}$ from DCE images
%Furthermore, the 5-fold cross-validation technique is used to find the best combination of input channels, the number of filters, and the size of filters for convolutional layers.

The use of 3D-CNN models has gained widespread popularity for classifying PCa based on volumetric image data due to their excellent performance. 
% A recent grand challenge, ``ProstateX'' \citep{armato2018prostatex}, shows that deep learning models have the ability to classify PCa by achieving state-of-the-art performance. 
\cite{mehrtash2017classification} propose a feature fusion 3D-CNN to classify clinically significant PCa using mp-MRI data. They use ADC maps, DWI, and K$^{trans}$ 3.0T MR data to enable the model to learn multi-modal information. Inspired by the VGG architecture \citep{simonyan2014very}, the model has three VGG-like feature extractors for each image modality, followed by the concatenation between outputs of each extractor and a vector represents the zonal information of the suspicious region. On the test set, the proposed model achieves the area under the receiver operating characteristic (AUC) curve of 0.80 on 140 unseen patients. \cite{liu2017prostate} propose a similar VGG-like 3D-CNN architecture for the same PCa classification task. Different from \cite{mehrtash2017classification}, they only have one model for feature extraction. To obtain the multi-modal information, they stack three images from each of the ADC maps, DWI, K$^{trans}$ into one 3-channel image as the input. The model achieves the AUC of 0.84 on the test set.      

In \cite{yoo2019prostate}, a probabilistic approach using mp-MRI data is employed for PCa classification. The authors develop an automated pipeline for the classification of clinically significant PCa using 3.0T DWI images from 427 patients. The pipeline consists of three parts: classification of each DWI slice using the pre-activated ResNet model \citep{he2016identity}, extraction and selection of first-order statistics from the CNN outputs, and final class label prediction using a random forest classifier. On the test set, the model achieves an AUC of 0.87. While the aforementioned studies may yield favorable AUCs, the reproducibility of the model might be challenging in clinics with limited patient (data) throughput. Recently, \cite{grebenisan2021spatial, grebenisan2020towards} address the data-hungry problem by introducing a disentangled representation learning approach (SDNet) to synthesize public 3.0T MRI images into 1.5T MRI images to increase the training data size for centres with limited 1.5T data. %Taking inspiration from \cite{chartsias2019disentangled}, 
Their approach aims to separate the anatomy- and modality-specific features present in images, subsequently merging the 1.5T modality features with the 3.0T anatomical features to generate MRI images resembling those acquired at 1.5T. Finally, a simple 3D-CNN classifier is used for the binary classification of clinically significant PCa. The model outperforms the state-of-the-art performance in PCa classification through domain alignment between different data sources.

While these methods exhibit excellent classification performance, a common limitation is the absence of a confidence score for their predictions, which hinders their interpretability in clinical practice.

\subsection{Uncertainty Estimation}
%{\color{red} refer to ``Training deep neural networks with noisy clinical labels: toward accurate detection of prostate cancer in US data'', add high level concepts, more ref, less detail}
%{\color{red} add someting about co-teaching}
Recent studies in medical imaging have highlighted the detrimental impact of label noise on the performance of modern deep learning models \cite{karimi2020deep}. Conventional regularization techniques such as dropout, batch normalization, weight decay, etc. fail to effectively address this challenge \cite{arpit2017closer, zhang2021understanding}. Methods proposed to mitigate such problem can be broadly categorized into three groups \cite{song2022learning}: (1) Robust loss functions and loss adjustments \cite{van2015learning, charoenphakdee2019symmetric, zhang2018generalized} aiming to stabilize the model performance when optimizing its parameters; (2) Sample selection \cite{jiang2018mentornet, malach2017decoupling, wang2020training} aiming to select a subset of ``clean'' data from a batch of samples to compute the loss; and (3) Robust architectures \cite{han2018co, yu2019does} aiming to learn the same data by training multiple models with different initialization assess output stability. While these methods inherently handle the noisy label problem, they can not provide explicit uncertainty estimation in terms of confidence in their output. Moreover, the ability of deep learning models to identify irrelevant samples remains limited. For instance, when a model trained on prostate MRIs is presented with a CT scan of the prostate at the time of inference, it is unclear whether the model can provide meaningful predictions or simply indicate a lack of in-domain knowledge and perform a human-in-the-loop analysis instead. In recent years, research has been conducted on uncertainty estimation for deep learning models. \cite{gal2016dropout, Gal2015} develop the \textit{dropout neural networks} framework to represent the prediction uncertainty of deep learning models, where the dropout layers in the model are formed by Bernoulli distributed random variables. During the test phase, predictive uncertainty is determined by enabling dropout layers and averaging the results over multiple runs, providing a valuable mechanism for uncertainty quantification in model predictions.
%To estimate the prediction uncertainty, one could simply collect results from several stochastic forward passes \citep{gal2016dropout} through the model. %Unlike Bayesian neural networks that implicitly model the prediction uncertainty through weight uncertainties, 
An alternative method for modeling uncertainty in deep learning models is through the use of \textit{evidential neural networks} \cite{sensoy2018evidential}, which formulate uncertainty by fitting a Dirichlet distribution - acting as the conjugate prior of the categorical distribution - to the class probabilities acquired from neural networks. This method considers model predictions as multinomial subjective opinions \citep{Joesang2016} or beliefs \citep{Dempster1968}, which can be further modeled explicitly using subjective logic. The "evidential" approach emphasizes the ability of the model to deliver certain predictions and exhibits superiority compared to the dropout approach \citep{gal2016dropout}.
%interpreted from the Dempster-Shafer Theory of Evidence \citep{Yager2008} perspective. %They modify three common loss functions according to the Dirichlet distribution: log-likelihood, cross-entropy, and the sum of squares loss. The Kullback-Leibler (KL) divergence loss is also utilized, and an annealing factor is used to control the weights for the KL loss. Experiments show that their method over-performs the MC Dropout method proposed by \cite{Gal2015}.  indispensable

In clinical practice, uncertainty estimation is crucial. By integrating uncertainty information into prediction outcomes, misclassification rates can be significantly reduced. For instance, in radiograph classification task \citep{Ghesu2019}, the authors employ the Dempster-Shafer Theory of Evidence \citep{Dempster1968} and the principles of subjective logic \citep{Joesang2016} to develop a framework that jointly estimates per-class probabilities and provides predictive uncertainty. This approach has been extended to abdominal ultrasound and brain MR images \citep{ghesu2021quantifying}. In the context of breast cancer classification, \cite{tardy2019uncertainty} apply the evidential neural networks approach \citep{sensoy2018evidential} to effectively diagnose breast cancer. A similar approach is used for the same task by \cite{Yuan2020} through the evidence adjustment technique, which focuses on the difference in the risks of uncertain samples from different classes. Consequently, we build upon the work from \cite{sensoy2018evidential} by adding uncertainty estimation to improve the robustness of the model and the interpretability of predictions.

\section{Materials}
\subsection{Data}
In this work, we use both large publicly available ProstateX data and small private local clinical data. A visualization of sample images from both datasets is presented in Figure \ref{sample_dat}.

\noindent \textbf{ProstateX Grand Challenge Data (3.0T)}. The 3.0T data is provided by the International Society of Optics and Photonics in the ``ProstateX'' challenge \citep{litjens2017prostatex}.
% The goal of this competition is to develope deep learning models to classify clinically significant PCa.
The dataset contains T2-weighted (T2), maximum b-value diffusion, diffusion-weighted imaging (DWI) with apparent diffusion coefficient (ADC) maps, and $K^{trans}$ images of 346 patients undergoing prostate biopsies. T2 images show the anatomical structure of the prostate, and both the ADC maps and K$^{trans}$ could further highlight the differences between normal and abnormal (cancer) cells in the MRI scans \citep{kasivisvanathan2018mri, kasson2018imaging}. We only use 204 of the total 346 patients in this work since these are reserved as training data, and hence they are provided with the spatial location of the suspicious finding, and a binary label indicating whether or not there is cancer. The remaining 142 patients are reserved as the test set and no labels are provided, hence, we exclude those from our work.

% The rest of the patients are used for testing in the competition and hence with no labels.

\noindent \textbf{Kingston Health Science Center Data (1.5T)}. The local 1.5T data is obtained from the Kingston Health Science Center (KHSC), which contains 104 patients with the corresponding biopsy-confirmed cancer and the Gleason Score. For the local data, only T2, ADC, and b-value images are available. All patients MRI have the spatial location of the suspicious finding(s), the Gleason Score, and the binary label indicating whether it is a cancer lesion or not.

Since all patients in both datasets have complete T2 and ADC data, our focus in this work is solely on these two types of images. Each MRI data in our study is associated with a single patient. Both datasets are processed similarly unless stated.

% \begin{figure*}[ht]
% 	\begin{center}
% 		%\fbox{\rule{0pt}{2in} \rule{.9\linewidth}{0pt}}
% 		\includegraphics[width=.99\textwidth]{Figures/sample_dat_vis.png}
% 	\end{center}
% 	%\vspace{-3ex}
% 	\caption{Visualization of sample data. From left to right are the T2 and ADC images from our local dataset; T2 and ADC images from the ``ProstateX Challenge''.}
% 	\label{sample_dat}
% 	%\vspace{-3ex}
% \end{figure*}
\begin{figure}[!tb]
     \centering
     \begin{subfigure}[b]{0.24\textwidth}
         \centering
         \includegraphics[width=\textwidth]{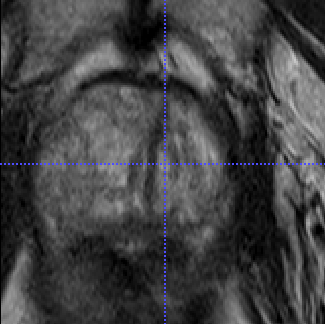}
         \caption{}
         \label{fig:t2kgh}
     \end{subfigure}
     \hfill
     \begin{subfigure}[b]{0.24\textwidth}
         \centering
         \includegraphics[width=\textwidth]{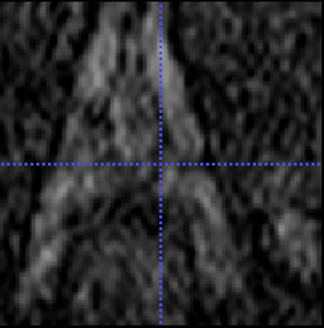}
         \caption{}
         \label{fig:adckgh}
     \end{subfigure}
     \hfill
     \begin{subfigure}[b]{0.24\textwidth}
         \centering
         \includegraphics[width=\textwidth]{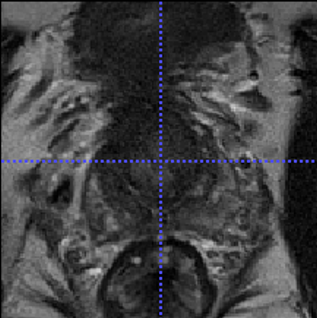}
         \caption{}
         \label{fig:t2px}
     \end{subfigure}
     \hfill
     \begin{subfigure}[b]{0.24\textwidth}
         \centering
         \includegraphics[width=\textwidth]{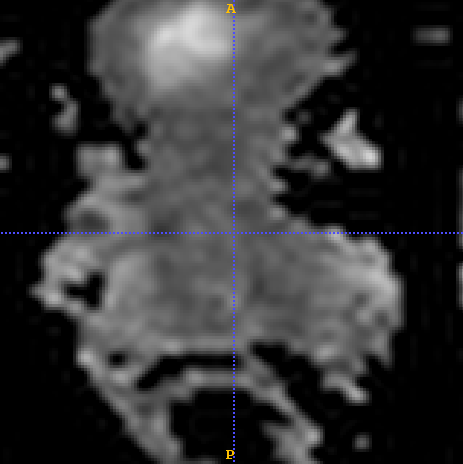}
         \caption{}
         \label{fig:adcpx}
     \end{subfigure}
     \caption{Visualization of sample data. \ref{fig:t2kgh} and \ref{fig:adckgh} are the 1.5T T2 and ADC images from KHSC, respectively. Similarly, \ref{fig:t2px} and \ref{fig:adcpx} are the 3.0T T2 and ADC images from the ``ProstateX'' Challenge, respectively.}
     \label{sample_dat}
\end{figure}

\subsection{Pre-processing} \label{data_prep}
 
 T2 and ADC sequences from both datasets are $160 \times 160 \times C$, where $C$ is the total number of slices in the MRI. We resample all 3D data to have the same voxel spacing. To reduce aliasing artifacts, the most common voxel spacing ($0.5 \times 0.5 \times 3~mm^3$) is used across all data, and the consine-windowed interpolation is utilized during sampling. We normalize pixel intensities to $[-1,1]$ for all data. For the translation purpose from 3.0T to 1.5T, we further resample all 3D data to $256 \times 256 \times C$ and split into $C$ 2D gray-scale slices. 
 % For the translation model, We use 90\% of the data from both datasets and perform validation on the remaining data.
 % % To train our GAN-based model, we consider all patients from both datasets. We use 90\% of the data from both datasets and perform validation on the remaining data.
 
 % For our classification model, the data cohort contains 204 patients from ProstateX dataset that already translated to 1.5T, and 104 patients in 1.5T from our local hospital. We use all data from ProstateX and 70 patients from our local hospital for training. The rest of 34 patients in our local hospital are used as test data. We split the training data into 80\% for actual training and 20\% for validation.
 % The augmented data is only used in classification models.
 \textbf{Augmentation: }For each patient, the MRI volume undergoes rotation ranging from 0 to 100 degrees in 5-degree increments, hence expanding the data size 20-fold.
 
 \textbf{Cropped Patches: }To reduce the computational cost, cropped patches of the MRI volume were employed. The process involves identifying the suspicious slice ($i_s$) based on the provided spatial location. Recognizing that PCa lesions can span multiple slices, two neighboring slices ($i_{s-1}$ and $i_{s+1}$) are selected as well and cropped around the biopsy location to generate a patch of size $64 \times 64 \times 3$.
% A methodological, model, or similar section often comes here.
 
\section{Methods} \label{all_methods}

Figure \ref{dtsche} summarizes an overview of our proposed approach. The domain transfer framework aims to reduce the distribution-level discrepancy between two prostate MRI datasets. The framework matches the acquisition parameters of publicly available, large 3.0T prostate mp-MRI data with local, small 1.5T prostate mp-MRI data. Once all the data from 3.0T are translated to 1.5T, a subsequent classifier is trained to classify clinically significant PCa. Furthermore, during the training process, the uncertainty is calculated along with the class output. We also introduce a novel evidential focal loss for the PCa classification task. Lastly, we utilize dataset filtering to improve robustness and accuracy by eliminating uncertain data samples from the training set.

\begin{figure*}[h!] % position of the figure (h)
	\begin{center}
		\includegraphics[width=.99\textwidth]{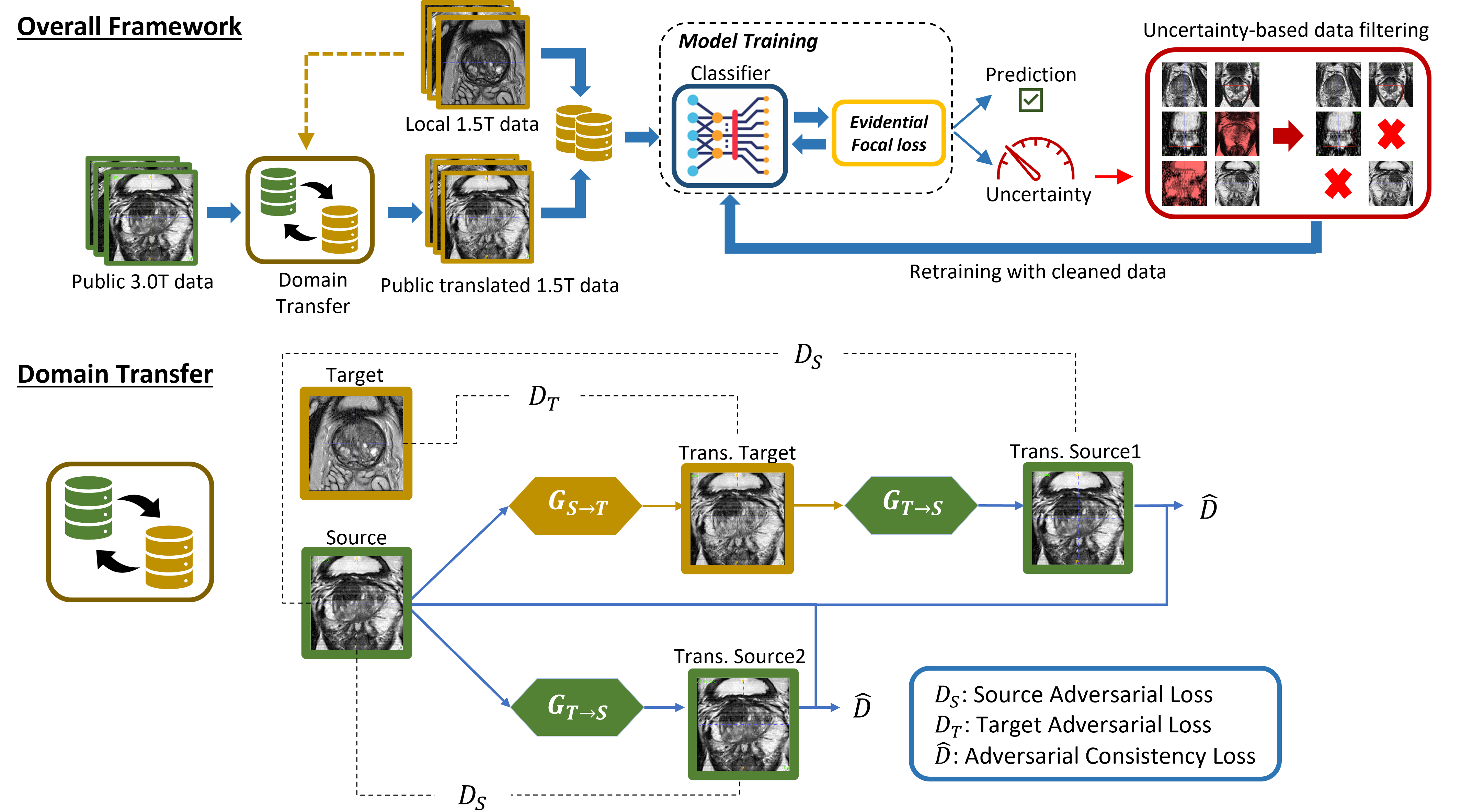}
	\end{center}
	%\vspace{-3ex}
	\caption{Detailed schematic of the proposed method. The overall framework of our proposed method contains two stages: 1), domain translation to map public 3.0T MRI with local 1.5T MRI; 2), uncertainty-aware clinically significant PCa classification. The bottom figure is the training schema for domain transfer. The upper right portion of the figure illustrates the PCa classification training process, which involves training the classifier using the Evidential Focal loss, filtering the training set based on uncertainty, and retraining the classifier on the filtered data to obtain the final classifier.}
	\label{dtsche}
	%\vspace{-3ex}
\end{figure*}

%% then add part of this portion for the rest of the structure
% We first provide an overview of our method that address the problem we mentioned in the previous section. We then describe the domain transfer framework and the choice of hyper-parameters. We discuss the uncertainty estimation for our method and the derivation of the \textit{Evidential Focal Loss}. Lastly, we detailed model architecture, training with filtration, and the evaluation process.

% \subsection{Method Overview}
% As we summarized in Figure \ref{dtsche}, the domain transfer framework could match the acquisition parameters of publicly available, enormous 3.0T prostate mp-MRI data with the local, small 1.5T prostate mp-MRI data. Once all the data from 3.0T are translated to the 1.5T-like data, the classifier is trained on these data. Further, during the training process, the uncertainty information is calculated along with the class output. We also utilize the filtration method to improve robustness and accuracy by eliminating the high predictive uncertainty data samples in the training set.

\subsection{The Domain Transfer Framework} \label{sec:dt}
%The goal of our domain transfer framework is to increase the number of 1.5T MRI training samples. It is a process to translate 3.0T prostate MRIs to 1.5T prostate MRIs by replacing the modality features in 3.0T with those in 1.5T, while maintaining the anatomy features in 3.0T. 
We adopt the ACL-GAN model \cite{zhao2020unpaired} for unpaired MR image translation from 3.0T to 1.5T. We have made modifications to the architecture to adapt it to single-channel MRI slices. There are two generators in this model namely $G_{T->S}$ and $G_{S->T}$, $G_{T->S}$ translates the images from the target domain ($X_T$) to the source domain ($X_S$) given the input $x \in X_T$ and a noise vector $z$ sampled from $\mathcal{N}(0,1)$. On the other hand, $G_{S->T}$ performs the reverse process of $G_{T->S}$, translating images from the source domain to the target domain. There are three discriminators, $D_S, D_T$, and $\hat{D}$ in this model. The first two ensure that translated images are in their respective domains by optimizing adversarial losses. The third discriminator, $\hat{D}$, ensures that translated images retain anatomical features in 3.0T by distinguishing the pair (Source, Trans. Source1) and (Source, Trans. Source2), as shown in the bottom of Figure \ref{dtsche}.
%This framework is a similar to the one proposed by \cite{grebenisan2021spatial}. However, the difference is we reconstruct the targeted 1.5T-like unpaired images by generative model through an adversarial manner.
The loss function of ACL-GAN contains four parts \cite{zhao2020unpaired}. First, the adversarial $\mathcal{L}_{adv}$ loss aims to encourage the translated image, either from source to target domain or from target to source domain, to be in the correct domain. The $\mathcal{L}_{adv}$ is further decomposed to source-domain and target-domain adversarial loss, $\mathcal{L}^S_{adv}$ and $\mathcal{L}^T_{adv}$, respectively. The $\mathcal{L}^T_{adv}$ is given by: 
\begin{equation}
\begin{aligned}
\mathcal{L}^T_{adv}(G_{S->T}, D_T, X_S, X_T) &= \mathbb{E}_{x_T \sim p_{T}} [log D_T(x_T)] 
\\&
+\mathbb{E}_{\bar{x}_T \sim p_{\{\bar{x}_T\}}}[log (1-D_T(\bar{x}_T))]
\end{aligned}
\label{equi:loss_adv_T}
\end{equation}
where $\bar{x}_T = G_{S->T}(x_S, z_1)$ and $z_1\sim \mathcal{N}(0,1)$. Similarly, the $\mathcal{L}^S_{adv}$ is given by: 
\begin{equation}
\begin{aligned}
\mathcal{L}^S_{adv}(G_{T->S}, D_S, \{\bar{x}_T\}, X_S) &= \mathbb{E}_{x_S \sim p_S} [log D_S(x_S)] \\
&+(\mathbb{E}_{\hat{x}_S \sim p_{\{\hat{x}_S\}}}[log (1-D_S(\hat{x}_S))] \\
&+\mathbb{E}_{\tilde{x}_S \sim p_{\{\tilde{x}_S\}}}[log (1-D_S(\tilde{x}_S))])/2
\end{aligned}
\label{equi:loss_adv_S}
\end{equation}
where $\hat{x}_S = G_{T->S}(\bar{x}_T, z_2)$, $\tilde{x}_S = G_{T->S}(x_S, z_3), z_2, z_3\sim \mathcal{N}(0,1)$. Combine these two adversarial losses we get the total adversarial loss for ACL-GAN, $\mathcal{L}_{adv} = \mathcal{L}^T_{adv} + \mathcal{L}^S_{adv}$.

A problem of $\mathcal{L}_{adv}$ is that this loss can not encourage the translated image to the target domain $\bar{x}_T$ is similar to the corresponding source domain image $x_S$, as we do not want the model to change the anatomical structure or features as we discussed previously. Hence, the adversarial consistency loss is proposed, which is given by:
\begin{equation}
\begin{aligned}
\mathcal{L}_{acl} &= \mathbb{E}_{(x_S, \hat{x}_S) \sim p_{(X_S, \{\hat{x}_S\})}}[log \hat{D}(x_S, \hat{x}_S)]\\
&+ \mathbb{E}_{(x_S, \tilde{x}_S) \sim p_{(X_S, \{\tilde{x}_S\})}}[log (1 - \hat{D}(x_S, \tilde{x}_S))]
\end{aligned}
\label{equi:loss_acl}
\end{equation}
where $\bar{x}_T=G_{S->T}(x_S, z_1)$, $\hat{x}_S=G_{T->S}(\bar{x}_T, z_2)$, $\tilde{x}_S=G_{T->S}(x_S, z_3)$.

Next, $\mathcal{L}_{idt}$ is the identity loss, which encourages generators to perform approximately identity mapping when images in the respective domain are provided, e.g., $x_S$ to $G_{T->S}$ or $x_T$ to $G_{S->T}$. The $\mathcal{L}_{idt}$ is given by:
\begin{equation}
\begin{aligned}
\mathcal{L}_{idt} = \mathbb{E}_{x_S \sim p_S} [|| x_S - x_S^{idt} ||_1] + \mathbb{E}_{x_T \sim p_T} [|| x_T - x_T^{idt} ||_1]
\end{aligned}
\label{equi:loss_idt}
\end{equation}
where $E^z_S: X_S \rightarrow Z$ and $E^z_T: X_T \rightarrow Z$ are two noise encoder networks for $G_S$ and $G_T$, respectively, which map images to noise vectors. $x_S^{idt} = G_{T->S}(x_S, E^z_S(x_S))$ and $x_T^{idt} = G_{S->T}(x_T, E^z_T(x_T))$.

Finally, $\mathcal{L}_{mask}$ is used to force both generators to only modify certain regions of the source image and keep the rest of the areas unchanged. We let generators produce a two-channel image, where the first channel is one of the translated images between the source and target domain (i.e., one-channel gray-scale prostate MRIs), and the second channel is the bounded mask, whose values are between $[0,1]$. The $\mathcal{L}_{mask}$ is given by:
\begin{equation}
\begin{aligned}
\mathcal{L}_{mask} &= \delta[(\max \{\sum_k x_m[k] - \delta_{max}\times P_t, 0\})^2\\
& + (\max \{\delta_{min}\times P_t - \sum_k x_m[k], 0\})^2]\\
& + \sum_k \frac{1}{|x_m[k]-0.5|+\epsilon}
\end{aligned}
\label{equi:loss_focus}
\end{equation}

where $\delta$, $\delta_{max}$ and $\delta_{min}$ are hyper-parameters for controlling the size of masks, $x_m[k]$ is the k-th pixel of the mask and $P_t$ is the total number of pixels in an image. The $\epsilon$ is a very small value to avoid dividing by zero. The first term of this loss controls the size of the mask. It encourages the generator to perform sufficient modifications while preserving the background information. Here, $\delta_{max}$ and $\delta_{min}$ are the maximum and minimum proportions of the foreground in the mask, i.e., the region we want to modify. The last term of this loss encourages the mask to be binary, either 0 or 1 to segment the foreground and background of the input image.  

Aggregating all loss terms together, we have:
\begin{equation}
    \mathcal{L}_{total} = \mathcal{L}_{adv} + \lambda_{acl}\mathcal{L}_{acl} + \lambda_{idt}\mathcal{L}_{idt} + \lambda_{mask}\mathcal{L}_{mask}
    \label{acl_loss}
\end{equation}
where $\lambda_{acl}, \lambda_{idt}, \lambda_{mask}$ are the weighting factors for $\mathcal{L}_{acl}, \mathcal{L}_{idt}, \mathcal{L}_{mask}$, respectively.

\subsection{Uncertainty-aware PCa Classification} \label{model_arc} % uncertainty-aware PCa classification
\subsubsection{Classifier architectures} \label{model_archis}
The traditional CNN approach is used for the clinically significant PCa binary classification task. Specifically, we explore three different model architectures for combinations of T2 and ADC patches as the classifier (Top row of Figure \ref{dtsche}). The first architecture, called the multi-stream CNN (M.S. MpMRI), treats T2 and ADC patches as separate inputs, as shown in Figure \ref{mulhead}. The model takes 3D patches of T2 and ADC as parallel inputs, which are then processed by the same feature extractor (i.e., weights are shared) to extract deep semantic representations. The output representations of T2 and ADC are then concatenated channel-wise and fed into another convolutional layer followed by a fully connected layer to produce the class probabilities.

%Recall that in Section \ref{data_prep}, we locate the suspicious slice $i_s$ and take two neighborhood slices $i_{s+1}$ and $i_{s-1}$ to form 3D patches. Hence,
In the second architecture, we adopt two different ways to combine ADC and T2 patches into a single input for the network. First, we stack cropped 3D patches of T2 and ADC along the channel axis, resulting in an input data size of $64 \times 64 \times 6$. Alternatively, we consider only the located suspicious slice $i_s$ for both T2 and ADC, and stack them along the channel axis to obtain the input data size of $64 \times 64 \times 2$. The model architecture for both combinations is similar to Figure \ref{mulhead}, where there is only one branch and no concatenation afterward. We refer to the model with input size of $64 \times 64 \times 6$ (resp. input size $64 \times 64 \times 2$) as Vol. MpMRI (resp. MpMRI).
%We use 3D convolution as the first layer to enable the extraction of features from the $x,y,z$ plane, after that all convolutional layers are in 2D. Mixing 2D and 3D convolutions could reduce the number of parameters in the model while enabling feature extraction in 3D.

Lastly, we use only 3D T2 patches as input to match with the previous work \citep{grebenisan2021spatial}. The model architecture is as same as the one for MpMRI, and we refer to this model as ``T2-only''.

\begin{figure*}[h!]
	\begin{center}
		%\fbox{\rule{0pt}{2in} \rule{.9\linewidth}{0pt}}
		\includegraphics[width=.99\textwidth]{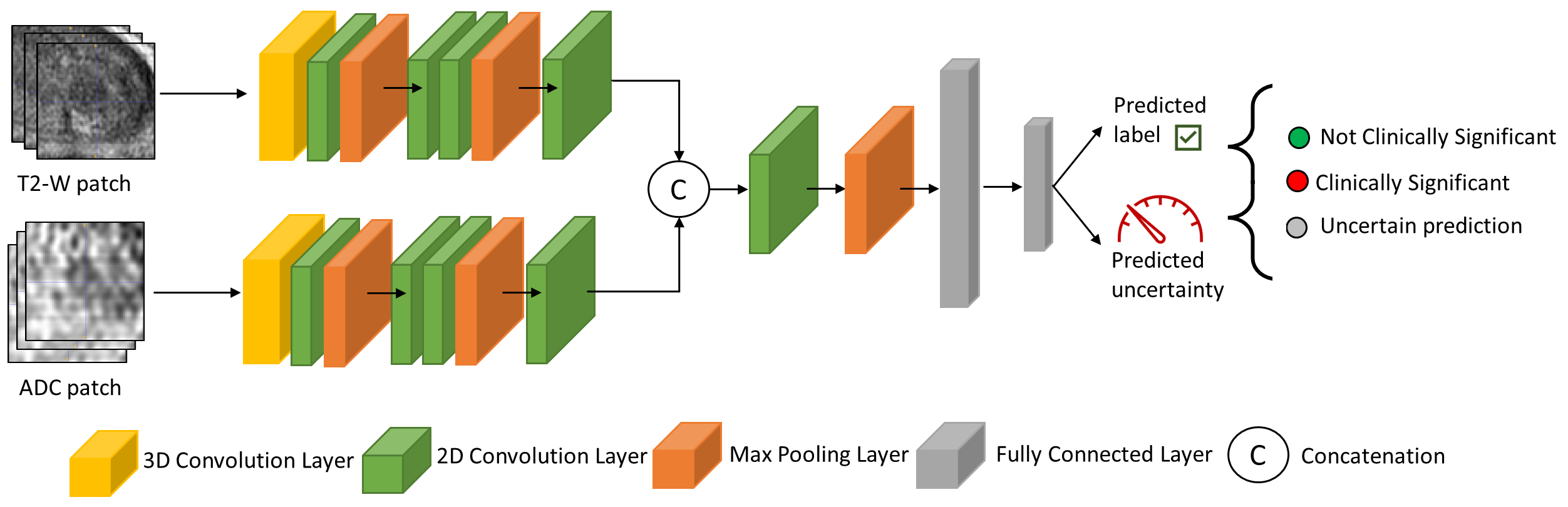}
	\end{center}
	%\vspace{-3ex}
	\caption{Detailed architecture of ``M.S. MpMRI'' model. The first sequence of CNN layers contains 1 $\times$ 3D convolution layer and 4 $\times$ 2D convolution layers, 2 $\times$ Max Pooling layers with window size $2 \times 2$. Both extracted feature maps of T2 and ADC are concatenated channel-wise. After that, another set of convolution-max pooling layers is utilized. Finally, the extracted 2D features are reshaped to 1D and fed into a Fully connected layer follow by a softmax layer with 2 outputs representing the probabilities of which class the input data belongs to.}
	\label{mulhead}
	%\vspace{-3ex}
\end{figure*}

We also explore with the \textit{co-teaching} framework \citep{han2018co} to combat the potential noisy label issue to improve the classification performance. This framework can be directly used as our classifier in a plug-and-play manner. The key distinction between the co-teaching framework and the conventional training paradigm lies in the fact that co-teaching only influences the number of samples in the batch that contribute to the loss calculation. We follow the exact setup in the original paper, and detailed hyperparameter settings can be found in Section \ref{exp_set}. The backbone model we selected for this framework is the ``MpMRI'' architecture, which we depicted previously.

\subsubsection{Evidential Focal Loss} \label{evi_focal}

Dataset filtering during the training phase could reduce the effect of the noisy label on the deep model. Followed by \cite{Ghesu2019}, the process of uncertainty-based filtering is shown at the top of Figure \ref{dtsche}: Firstly, we calculate the uncertainty value for each sample in the training set. We then remove a portion of the training samples that exhibit high predictive uncertainty. Finally, we retrain the model using the remaining ``clean'' training data.

% We define the above two filtering methods as follows:

% \noindent \textbf{Patch-driven filtering: }Given the training uncertainty for each data sample, we simply eliminated $x\%, x \in [10, 20]$ of training \textit{patches} with high uncertainty value and retrained on the rest of the samples on the training set.

% \noindent \textbf{Patient-driven filtering: }Given the training uncertainty for each data sample, recall that each patient associated with 20 patches that we discussed in Section\ref{data_prep}. To compute the uncertainty of each patient, we averaged the uncertainty value for the corresponding 20 patches. We then eliminated $x\%, x \in [10, 20]$ of the training \textit{patients} with high uncertainty value and retrained the rest on the training set.
% {\color{red} patch based, patient based filter}

%\subsubsection{Evidential Focal Loss}

Following the previous work \citep{sensoy2018evidential}, we extend and combine the idea of subjective logic \citep{Joesang2016} with the focal loss \citep{lin2017focal} for the clinically significant PCa binary classification task. In the context of the Theory of Evidence, a belief mass is assigned to individual attributes, e.g., the possible class label of a specific data sample. The belief mass is generally calculated from the evidence collected from the observed data \citep{Dempster1968}. Let $K$ be the number of classes, and $b_k \geq 0, k \in [1,K]$ is the belief mass for class $k$ and $u \geq 0$ is the overall uncertainty measure. Let $e_k \geq 0$ be the evidence computed for $k^{th}$ class, then the belief $b_k$ and the uncertainty $u$ are computed as follows:

\begin{equation} \label{belief_uncern}
    b_k = \frac{e_k}{S} \quad and \quad u = \frac{K}{S}
\end{equation}

where $S = \sum_{i=1}^{K} (e_i+1)$. For our binary task ($K=2$), we can further simplify Equation \eqref{belief_uncern} to $b_0 = \frac{e_0}{e_0+e_1+2}, b_1 = \frac{e_1}{e_0+e_1+2}$, and $u = \frac{2}{e_0+e_1+2}$. The belief mass assignment, e.g., subjective opinion, corresponds to the Dirichlet distribution with parameters $\alpha_{k} = e_k+1$, and $S = \sum_{i=1}^{K} \alpha_{k}$ is the Dirichlet strength. The expected probability for the $k^{th}$ class is calculated by the mean with the associate Dirichlet distribution, e.g., $\hat{p}_k = \frac{\alpha_k}{S}, k \in [1,...,K]$ \citep{sensoy2018evidential}.

% Formally, the Dirichlet distribution is a probability density function that models the possible values of the probability mass function $\mathbf{p}$. It is a multivariate generalization of beta distribution and is often used as the prior distribution in Bayesian statistics. Dirichlet distribution takes a vector-valued $\boldsymbol{\alpha}:=[\alpha_1, ..., \alpha_K]$ of length $K$ that has the same characteristics as the multinomial as input:

% $$ \mathbf{D}(\mathbf{p}|\boldsymbol{\alpha})=\left\{
% \begin{aligned}
% & \frac{1}{\mathcal{B(\alpha)}} \prod_{i=1}^{K} p_{i}^{\alpha_{i}-1}, & \ for \ \mathbf{p} \in S_K\\
% & 0, & otherwise
% \end{aligned}
% \right.
% $$

% where $\mathcal{B(\alpha)}$ is the $K$-dimensional multinomial beta function \citep{Kotz2004}, and $S_K$ is the $K$-dimensional unit simplex defined as:

% \begin{equation} \label{sk_simplex}
%     S_K = \{\mathbf{p}|\sum_{i=1}^{K} p_i = 1 \ and \ 0 \leq p_1,\ldots, p_K \leq 1\}
% \end{equation}

% Following \cite{sensoy2018evidential}, the expected probability for the $k^{th}$ class is calculated by the mean with the associate Dirichlet distribution:

% \begin{equation} \label{exp_prob}
%     \hat{p}_k = \frac{\alpha_k}{S}, k \in [1,...,K]
% \end{equation}

Now, we formally define our proposed evidential focal loss. Given the training set contains $N$ data samples, $D := \{x_i, y_i\}_{i=1}^{N}$, where $x_i$ is the $i^{th}$ data sample and $y_i \in [0,1]$ is the corresponding label, 0 is the negative sample and 1 is the positive sample. We further denote $\mathbf{y_i}$ as the one-hot encoding label for sample $i$, e.g., $\mathbf{y_i} = [1,0]$ for class 0 and $\mathbf{y_i} = [0,1]$ for class 1. The focal loss \citep{lin2017focal} for binary classification is defined by $FL(p_t) = -w_{t}(1-p_t)^{\gamma}\log(p_t)$ where $p_t = p$ if $y_i = 1$ for $i^{th}$ sample, otherwise $p_t = 1-p$ with probability output $p$ from the model, and $w$ is the class weight. Let $\mathbf{P}_{i}$ be a vector that contains the probability of $i^{th}$ sample for both classes from our model output; $p_{ij}$ is the probability of $i^{th}$ sample belonging to $j^{th}$ class; $K$ is the number of classes, and $\beta_j$ is the class weight of $j^{th}$ class. $\gamma$ is the focusing parameter to reduce the loss for well-classified samples, and we fix $\gamma = 2$ in this task. We could define Evidential Focal Loss as the following:

\begin{align} \label{efl_orig}
    \mathcal{L}^{cls}_{i}(\theta) = \int \sum_{j=1}^{K} -\beta_j(1-p_t)^\gamma \log(p_{ij}) \frac{1}{\mathcal{B}(\alpha_{i})} \prod_{j=1}^{K} p_{ij}^{\alpha_{ij}-1} d\mathbf{P}_{i}
\end{align}

Here, we denote $\mathcal{L}^{cls}_{i}(\theta)$ as the classification loss for a single sample $i$, $\alpha_i$ is the belief mass $\alpha$ for $i^{th}$ sample for all classes and $\alpha_{ij}$ is the belief mass $\alpha$ for $i^{th}$ sample and $j^{th}$ class. $\mathcal{B}(\cdot)$ is the multinomial beta function and $\frac{1}{\mathcal{B}(\alpha_{i})} \prod_{j=1}^{K} p_{ij}^{\alpha_{ij}-1}$ is the probability density function of Dirichlet distribution. Rewriting class probabilities in vector form, the equation \eqref{efl_orig} can be simplified to \eqref{efl_second} by the definition of expectations:

\begin{align} \label{efl_second}
    \mathcal{L}^{cls}_{i}(\theta) = - \sum_{j=1}^{K} \beta_j \mathbf{E}[(1 - \mathbf{P}_{i})^2 \log(p_{ij})]
\end{align}

Different from the original focal loss, we replace the constant term $1$ with $\mathbf{y_i}$, to tackle the hard-to-classified samples and reduce the loss of well-classified samples \textit{in both classes}. Recall that expected probability $\hat{p}_k$ for the $k^{th}$ class is $\alpha_{k} / S$, then by the linearity of expectations and the definition of expectations of Dirichlet distribution, we have:

\begin{align} \label{final_efl}
   \mathcal{L}^{cls}_{i}(\theta) = \sum_{j=1}^{K} {\beta_j} (y_{ij} - (\alpha_{j} / S))^2 (\psi(S_i) - \psi(\alpha_{ij})) %\mathcal{P}_{ij}
\end{align}

where $\psi(\cdot)$ is the digamma function, $y_{ij}$ is the $j^{th}$ class label in the one hot encoding representation $\mathbf{y_i}$ and $\beta_j$ is the class weight vector for class $j$.

To ensure that highly uncertain data samples, referred to as "I do not know" decisions, do not impact the overall data fit and to minimize their associated evidence, we adopt the Kullback-Leibler (KL) loss as the regularization term to penalize the unknown predictive distributions, as done in \citep{sensoy2018evidential}. Combining $\mathcal{L}^{cls}$ and the KL loss yields our final \textit{evidential focal loss} for uncertainty-aware classification: 
% This involves utilizing the Kullback-Leibler (KL) divergence as a regularization term to penalize the unknown predictive distributions, effectively shrinking their influence towards zero. The KL divergence is as same as it is defined in \cite{sensoy2018evidential}.
%as the following:

% \begin{align*} \label{kl_div}
%     KL[D(\mathbf{P}_{i}|\alpha_{i}) || D(\mathbf{P}_{i}|\mathbf{1})]
%     & = \log \Bigg ( \frac{\Gamma(\sum_{k=1}^K \tilde \alpha_{ik})}{\Gamma(K) \prod_{k=1}^K \Gamma(\tilde \alpha_{ik}) } \Bigg )+ \sum_{k=1}^K (\tilde \alpha_{ik}-1) \Bigg [\psi(\tilde \alpha_{ik})-\psi \Big (\sum_{j=1}^K \tilde \alpha_{ij} \Big ) \Bigg ]
% \end{align*}

% where $\mathbf{P}_i$ is still the class probabilities predicted by our model for $i^{th}$ sample, $D(\mathbf{P}_{i}|\mathbf{1})$ is the Dirichlet uniform distribution, $\mathbf{1}$ is a vector of $K$ ones, and $\tilde{\boldsymbol{\alpha_{i}}} = \mathbf{y}_i + (1-\mathbf{y}_i) \odot \boldsymbol{\alpha_{i}}$ is the Dirichlet distribution with the non-misleading evidence removed from the original predicted $\boldsymbol{\alpha_{i}}$ for $i^{th}$ sample.

\begin{align}
    \mathcal{L}^{efl}(\theta) = \sum_{i=1}^{N} \mathcal{L}^{cls}_{i}(\theta) + \lambda_t \sum_{i=1}^{N} KL[D(\mathbf{P}_{i}|\alpha_{i}) || D(\mathbf{P}_{i}|\mathbf{1})]
\end{align}

where $\mathbf{1}$ is an one-vector, $D(\mathbf{P}_{i}|\mathbf{1})$ is the uniform Dirichlet distribution. $\lambda_t$ is the weighting factor of the KL divergence loss and is defined as $\lambda_t = min(1.0, t/10) \in [0,1]$, where $t$ is the current number of epochs of training.

Finally, we introduce two proposed methods for filtering training samples based on the calculated uncertainty.

 \textbf{Patch-driven filtering: }Given the uncertainty for each training patch, we simply eliminate $x\%, x \in [10, 20]$ of the \textit{patches} with the highest uncertainty and retrain the model on the rest of the samples on the training set.

 \textbf{Patient-driven filtering: }Similar to Patch-driven filtering, we first calculate the uncertainty of each training patch. To determine the uncertainty of each patient, we calculate the average uncertainty value across their corresponding patches (20 patches per patient as mentioned in Section \ref{data_prep}). We then eliminate $x\%, x \in [10, 20]$ of the training \textit{patients} with high uncertainty value and retrain the model on the rest on the training set.

\section{Experiments \& Setup} \label{exp_set}
\subsection{Setup} \label{setup}

\textbf{Data Split: }For the domain transfer task, as mentioned in Section \ref{data_prep} and \ref{all_methods}, the resampled T2 and ADC ``images" with size $256 \times 256$ from both ProstateX and the local dataset is used for training the ACL-GAN model. Particularly, we allocate 90\% of images in both datasets as training and keep 10\% as the validation set to avoid overfitting the ACL-GAN model. Importantly, we ensure that the images corresponding to each patient are exclusively present in either the training or validation set, but not both. To improve the robustness and enhance the ability of the ACL-GAN to capture feature-level representations of 1.5T images, we use all data from our local hospital. However, it is important to note that this approach does not yield any additional impacts on the subsequent classification task. The model only modifies image regions that have visual differences caused by acquisition parameters of various MRI machines, but it does not alter the context of the prostate itself.

For the PCa classification task, we use cropped and augmented T2 and ADC \textit{patches} from both datasets. As mentioned before, this includes 204 ProstateX patients translated to 1.5T, as well as 104 patients from our local hospital captured in 1.5T. Regarding the data split for the classification, we keep patches of 34 patients from our local center as a standalone test set. From the remaining patches (70 local patients and all ProstateX patients), we allocate 80\% for training and 20\% for validation, assuring patches from the same patient are not included in both of these sets.

Next, we provide a brief intro to the experiments we conducted in this study.

\textbf{Domain Transfer: }The first experiment involved translating ProstateX MRI data from 3.0T to 1.5T using our proposed ACL-GAN model. We evaluated the effectiveness of our proposed approach by using quantitative metrics such as Fr\'echet Inception Distance (FID) score \cite{heusel2017gans} and maximum mean discrepancy (MMD) score \cite{gretton2012kernel}, more details could be found in Section \ref{sec:eval}. The translated MRI data was then employed in a downstream binary classification task for clinically significant PCa, demonstrating the superiority over the SDNet \cite{grebenisan2021spatial} baseline.

\textbf{Classification: }We divided our classification experiments into two folds. \textbf{(1). Conventional approach}, we used the conventional training paradigm without any filtering or uncertainty estimation. Three different model architectures were utilized for these experiments as discussed in Section \ref{model_archis}. \textbf{(2). Uncertainty-aware approach}, we used the dataset filtering method and evidential focal loss proposed in Section \ref{evi_focal} to train our models. Additionally, several ablation studies on data modalities, model architectures, and loss functions were conducted, with corresponding classification results detailed in Section \ref{res_dis}. Finally, we focused on dataset filtering during deployment and examined how this technique affects the classification performance on the test data.

%The rest of experiments are focused on the clinically significant PCa classification. We test our methods on 34 patients from our local hospital that are hidden from the training process and report the classification performance.

\subsection{Experimental Details}

We trained two ACL-GAN models separately for T2 and ADC images as part of our domain transfer framework. The optimizer used for both models was Stochastic Gradient Descent with Adam update rule \cite{kingma2014adam}, with an initial learning rate of 0.0001 and weight decay of 0.0001 to prevent overfitting. The batch size is 3 and are trained for 30,000 epochs. Moreover, when training the model for T2 images, we set the $\lambda_{mask} = 0.0025, \lambda_{idt} = 1, \lambda_{acl} = 0.2$ in Equation \eqref{acl_loss} and lower and upper mask threshold to be 0.005 and 0.1, respectively. When training the model for ADC images, the values of $\lambda_{mask}, \lambda_{idt}, \lambda_{acl}$ are the same as in the T2 model with lower and upper mask thresholds set to 0.001 and 0.005, respectively. We adopt the Least-Square (LS) loss \citep{mao2017least} for $\mathcal{L}_{adv}$ and $\mathcal{L}_{acl}$ in Equation \eqref{acl_loss}, as done in \cite{zhao2020unpaired}.

\textbf{Converting 3.0T to 1.5T: }Once we have obtained two ACL-GAN models, we need to standardize the acquisition parameters of 3.0T prostate MRIs to match those of the 1.5T data in our local dataset. To achieve this, we divided the original 3.0T MRI into multiple 2D grayscale slices. For each 2D slice, we used the generator $G_{T}$ and a noise vector $z$ randomly sampled from $\mathcal{N}(0,1)$ to translate the 3.0T slice to 1.5T, e.g., $I_{\text{1.5T}} = G_{T}(I_{\text{3.0T}}, z)$ as discussed in Section \ref{sec:dt}. This process was repeated for all 2D slices, and the slices were stacked back together to reconstruct the 3D MRI for each patient. The voxel spacing remained unchanged before and after the translation process. This procedure was applied to both T2 and ADC data.

All classification models were trained with Stochastic Gradient Descent with Adam, and batch normalization was applied to expedite convergence. In the first category of classification experiments (Conventional approach), the traditional focal loss \citep{lin2017focal} with $\gamma = 2$ was employed. Specifically, all models except co-teaching were trained for 300 epochs with a learning rate of 0.0001, weight decay of 0.01, and a batch size of 10. For the co-teaching model, the noise rate and forget rate were set to 0.1, and the number of epochs for the linear drop rate to 10. The model was trained for 300 epochs, with a batch size set to 10, and a learning rate of 0.00001. In the second category of experiments (Uncertainty-aware approach), a learning rate of 0.0001 was used and decayed by a factor of 0.1 every 200 epochs. The weight decay was set to 0.01, the total training epochs were 300, and the batch size was 10. Further training details can be found in Appendix \ref{app:exp_det}.

%The set of experiments on PCa classification without filtration and without model prediction uncertainty. We also showed the different combinations of T2 and ADC images as model input; different model architectures and performed the ablation study accordingly. Moreover, we designed a set of experiments on the uncertainty-aware PCa classification. The filtration method happened in either patch-driven or patient-driven. Finally, we focused on filtration while deployment, where we aimed to see how the filtration affected the classification performance.
%{\color{red} brief talk about experiments to be done, summary of experiments}

% \subsection{Converting Patch-based results to Patient-based results} \label{sec:p2p}
% To generate the results at the patient level, we first use our classifier to predict the test patches. The outputs from the classifier are separated into $x$ groups sequentially, where $x$ is the number of patients in the test dataset. Denote $P_i \in \mathbb{R}^2, i\in[1,...,x]$ be the vector contains probabilities for each patient $i$. For each $P_i, i \in [1,...,x]$, we compute the median value $\tilde{p}_{i}$ of the vector and use this value as the probability for $i$^{th} patient. Finally, a threshold of 0.5 is used to determine whether the patient has PCa or not, i.e., assign label 1 if $\tilde{p} > 0.5$, otherwise assign label 0.

\subsection{Evaluation} \label{sec:eval}

%{\color{red} talk about patch-patient first}
To access the quality of the translated images and validate the effectiveness of our proposed domain transfer framework for ProstateX data from 3.0T to 1.5T, we followed previous works \cite{kwon2019generation,volokitin2020modelling,zhou2024conditional} to compute the maximum mean discrepancy (MMD) score \cite{gretton2012kernel} and the Fr\'echet Inception Distance (FID) score \cite{heusel2017gans}. We followed the implementation provided in the Project-MONAI library \footnote{https://github.com/Project-MONAI/GenerativeModels/tree/main/generative/metrics}. Both MMD and FID score measure the distribution distance between translated 1.5T images from the ProstateX dataset and real 1.5T images from our local hospital. Lower values in these metrics indicate greater fidelity to the real data distribution. The translated 1.5T images' quality was further validated in a downstream classification task, combining them with the original 1.5T images to train a classifier distinguishing patients with and without clinically significant PCa. Traditional classification metrics, including accuracy (Acc.), sensitivity (Sen.), specificity (Spec.), and Area under ROC curve (AUC), were employed for this evaluation. Furthermore, we used ``uncertainty calibration'' to assess for performance of our uncertainty-based models. To compute calibration, we employed Expected Calibration Error (ECE) as done in \cite{gilany2022towards,mehrtash2020confidence}. ECE measures the correspondence between the confidence of predictions and the actual model accuracy. Unlike other classification metrics, smaller values of ECE indicate less miscalibration, hence indicating better model calibration capability.

Reporting the patient-level performance is more relevant to the real clinical setting. However, since patches were used as model input, aggregating individual results from these patches is necessary to calculate patient-level metrics. This involved using the classifier to predict test patches, which were then sequentially grouped into $x$ groups, where $x$ is the number of patients in the test dataset. For each patient, there are 20 patches hence probabilities due to the mentioned augmentation in Section \ref{data_prep}. The \textit{median} probability $\tilde{p}{i}$ was computed over 20 probabilities as the aggregated probability for each patient. Finally, a threshold of 0.5 was applied to determine whether a patient has PCa, assigning label 1 if $\tilde{p}{i} > 0.5$ and 0 otherwise.

% \begin{figure*}[h!]
% 	\begin{center}
% 		%\fbox{\rule{0pt}{2in} \rule{.9\linewidth}{0pt}}
% 		\includegraphics[width=.65\textwidth]{Figures/evaluation.png}
% 	\end{center}
% 	%\vspace{-3ex}
% 	\caption{The detailed evaluation process. The model could say "I do not know" for a specific data sample if its uncertainty value exceeds a pre-defined threshold.}
% 	\label{eval}
% 	%\vspace{-3ex}
% \end{figure*}

\section{Results and Discussion} \label{res_dis}

In this section, we analyze the translated image from our proposed domain transfer framework and report \textbf{patient-based} classification results; the performance of our methods on patches are reported in Appendix \ref{app:pat_res}.

\subsection{Quantitative Analysis of Translated Samples} \label{dt_vres}

% Figure \ref{qualres} shows a sample visualization of the difference between original 3.0T T2 and ADC images in the ProstateX Challenge and the corresponding translated 1.5T T2 and ADC images using our proposed domain transfer framework. As shown, domain transfer reduces the image contrast and results in the loss of minor details in the original 3.0T images.
We evaluate the quality of translated T2 images using two common metrics MMD \cite{gretton2012kernel} and FID \cite{heusel2017gans} score, as reported in Table \ref{quan_res}. The FID score is computed over each slice in the MRI data, i.e., a $160 \times 160 \times 32$ MRI volume results in 32 individual slices and FID scores, and we take the average scrore across all slices as the final score. The ACL-GAN model we used outperforms the SDNet in both metrics, we attribute the observed improvements to several key factors. Unlike SDNet \citep{grebenisan2021spatial}, which merges modality features from randomly selected 1.5T images with anatomical features from 3.0T, our method learns the overall data distribution in an adversarial manner. This enables it to capture the entire distribution of 1.5T images and perform the translation more effectively. Additionally, our approach ensures that the translated image retains crucial features of the original, with the generator making modifications only to specific parts of the image. These designs contribute to the superior performance of our ACL-GAN-based domain transfer framework.

% we argue that the improvements are based on the following reasons. In SDNet \citep{grebenisan2021spatial}, modality features from randomly selected 1.5T images are merged with anatomical features from 3.0T rather than from the whole distribution of all 1.5T images. In contrast, our method learns the overall data distribution in an adversarial manner, capturing the entire distribution of 1.5T images and performing the translation. Moreover, our method ensures the translated image contains the crucial features of the original image, and the generator only modifies certain parts of the image.

Since SDNet \cite{grebenisan2021spatial} is solely trained on T2 images, a direct quantitative comparison for translated ADC images is omitted. Instead, we conducted an ablation study to highlight the improvement in classification performance of adding translated ADC images to our model, see details in Section \ref{pca_res_nofil}.

\begin{table}[ht]
\centering
\begin{tabular}{c|c|c}
           & MMD $\downarrow$ & FID $\downarrow$  \\ \hline
SDNet \cite{grebenisan2021spatial} & 0.0018 & 12.740   \\
%3D-$\alpha$GAN & 46158 & 0.925\\
ACL-GAN (ours) & \textbf{0.00054} & \textbf{11.464} \\ \hline    
\end{tabular}
\caption{Quantitative results of translated T2 images using the baseline method SDNet and ACL-GAN in our proposed domain transfer framework. Lower values indicate better performance for all metrics. \textbf{Bold} values represent the best results.}
\label{quan_res}
\end{table}

\subsection{PCa classification without filtering} \label{pca_res_nofil}

Table \ref{pat_res} summarizes the experiment results in this section, which contains the PCa classification performance using the conventional approach, as detailed in Section \ref{setup}. We observed that the AUC of using the co-teaching framework with MpMRI architecture as the base model achieves the best AUC and outperforms the baseline. The co-teaching model exhibits approximately a 50\% increase in sensitivity while experiencing only a modest ~10\% decrease in specificity compared to the baseline model. This suggests that the co-teaching model demonstrates superior learning capabilities for classifying both positive and negative data samples on the test set. In the training process, we adopt a greedy approach of assuming 10\% of the samples to be noisy. Consequently, both models need to designate a portion of the data in each batch as ``clean" to update the parameters. This strategy allowed our model to prioritize learning from the clean data, leading to enhanced robustness.

\textbf{Ablation Study:} We embed the results of the ablation study in Table \ref{pat_res}, which explores two key variations: alteration of the number of input modalities and alteration of the architecture of the model. To assess the impact of data modalities on classification performance, the T2-only model is compared with MpMRI and M.S. MpMRI models, both use T2 and ADC patches as input. The addition of the ADC modality leads to a substantial improvement in classification performance, highlighting the utility of multi-modal information in guiding the model for clinically significant PCa classification. The examination of model architecture reveals that the model with simpler inputs, MpMRI, performs better. Furthermore, the results can be further enhanced by leveraging the co-teaching framework.

% \begin{figure}[H]
% \centering
% \includesvg[width=130mm]{Figures/result_comparison_new.svg}
% \caption{AUC curves of the selected experiments. The shaded areas represent the 95\% confidence intervals (CI) of each model. CIs are obtained by using bootstrap with $n = 1000$.}
% \label{pac_vis}
% \end{figure}

% \begin{table}[H]
% \centering
% \begin{tabular}{|c|c|c|c|c|c|}
% \hline
% \multicolumn{1}{|l|}{} & Data   & Acc. & Sen. & Spec. & AUC  \\ \hline
% SDNet(baseline)        & T2     & 79.4 & 28.6 & \textbf{92.6}  & 76.2 \\ \hline
% T2-only                & T2     & 64.7 & 71.4 & 63.0  & 77.8 \\ \hline
% MpMRI                  & T2+ADC & 79.4 & 85.7 & 77.8  & 84.7 \\ \hline
% Vol. MpMRI             & T2+ADC & 67.6 & 71.4 & 66.7  & 68.9 \\ \hline
% M.S. MpMRI             & T2+ADC & 73.5 & 71.4 & 74.1  & 82.5 \\ \hline
% MpMRI+co-teaching      & T2+ADC & \textbf{82.3} & \textbf{85.7} & 81.2  & \textbf{88.4} \\ \hline
% \end{tabular}
% \caption{\textbf{Patient-based results} of the performed experiment. Acc., Sen., Spec. and AUC are the shorts for Accuracy, Sensitivity, Specificity and Area Under (ROC)Curve, respectively. Best results are in \textbf{bold}. All units of the numeric values are in \%.}
% \label{pat_res}
% \end{table}

\begin{table}[!ht]
\centering
\begin{tabular}{c|c|c|c|c|cl}
                                       & \multicolumn{1}{c|}{Data}    & Acc.          & Sen.          & Spec.         & AUC           &  \\ \cline{1-6}
\multicolumn{1}{c|}{SDNet\cite{grebenisan2021spatial}(baseline)}   & \multicolumn{1}{c|}{T2}      & 79.4          & 28.6          & \textbf{92.6} & 76.2$\pm$17.5          &  \\
\multicolumn{1}{c|}{T2-only}           & \multicolumn{1}{c|}{T2}      & 64.7          & 71.4          & 63.0          & 77.8$\pm$18.0          &  \\ \hline
\multicolumn{1}{c|}{MpMRI}             & \multicolumn{1}{c|}{T2+ADC}    & 79.4          & \textbf{85.7} & 77.8          & 84.7$\pm$15.5          &  \\
\multicolumn{1}{c|}{Vol. MpMRI}        & \multicolumn{1}{c|}{T2+ADC}    & 67.6          & 71.4          & 66.7          & 68.9$\pm$26.1          &  \\
\multicolumn{1}{c|}{M.S. MpMRI}        & \multicolumn{1}{c|}{T2+ADC}    & 73.5          & 71.4          & 74.1          & 82.5$\pm$14.3          &  \\
\multicolumn{1}{c|}{MpMRI+co-teaching} & \multicolumn{1}{c|}{T2+ADC}    & \textbf{82.3} & \textbf{85.7} & 81.2          & \textbf{88.4$\pm$10.6} &  \\ \cline{1-6}
% \multicolumn{1}{c|}{MpMRI}             & T2+ADC                      & 10\% & patch   & 82.4          & 85.7          & 81.5          & 85.7$\pm$9.9          &  \\
% \multicolumn{1}{c|}{M.S.MpMRI}         & T2+ADC                      & 10\% & patch   & 82.4          & 71.4          & 85.2          & 83.6$\pm$13.5          &  \\
% \multicolumn{1}{c|}{MpMRI}             & T2+ADC                      & 20\% & patch   & \textbf{\color{blue}85.3}          & \textbf{\color{blue}100}           & 81.5          & \textbf{\color{blue}98.4$\pm$1.6}          &  \\
% \multicolumn{1}{c|}{M.S. MpMRI}        & T2+ADC                      & 20\% & patch   & \textbf{\color{blue}85.3}          & 71.4          & \textbf{\color{blue}88.9}          & 92.6$\pm$7.4          &  \\ \cline{1-8}
% \multicolumn{1}{c|}{MpMRI}             & T2+ADC                      & 10\% & patient & \textbf{\color{red}88.2}          & \textbf{\color{red}100}           & \textbf{\color{red}85.2}          & \textbf{\color{red}92.6$\pm$7.4}          &  \\
% \multicolumn{1}{c|}{M.S. MpMRI}        & T2+ADC                      & 10\% & patient & 85.3          & \textbf{\color{red}100}           & 81.5          & 86.2$\pm$9.0          &  \\
% \multicolumn{1}{c|}{MpMRI}             & T2+ADC                      & 20\% & patient & 73.5          & 85.7          & 70.4          & 86.8$\pm$12.6          &  \\
% \multicolumn{1}{c|}{M.S. MpMRI}        & T2+ADC                      & 20\% & patient & 73.5          & 71.4          & 74.1          & 84.6$\pm$14.2          &  \\ \cline{1-8}
\end{tabular}
\caption{\textbf{Patient-based results} of experiments using conventional training paradigm in Section \ref{setup}. Standard deviations are computed from the 95\% bootstrap confidence interval with $n = 3000$ samples. The best results are \textbf{bold}. All units of the numeric values are in \%.}
\label{pat_res}
\end{table}

\subsection{PCa classification with filtration} \label{pca_res_fil}

In this section, we embark on experiments utilizing two different architectures, MpMRI and M.S. MpMRI, and incorporating training set filtering at various rates. The evidential focal loss described in Section \ref{evi_focal} is used as the loss function to optimize the models. The co-teaching framework is excluded from this section for the following reason: while co-teaching \textit{implicitly} handles noisy labels or samples in the training set, the training set filtering in Section \ref{evi_focal} is an explicit alternative to dealing with them. The co-teaching framework will first update its model parameters with simpler and cleaner samples during training. However, through the filtering process, data samples with high uncertainty values are considered potentially noisy and hence are not involved in the training process. We argue that the use of co-teaching and simultaneous data filtering might be redundant. Our hypothesis for training set filtering is that by explicitly eliminating highly uncertain data samples from the set and optimizing only on the remaining "confident" samples using the evidential focal loss (Section \ref{evi_focal}), we can produce a more robust model. Therefore, to coalesce our proposed loss function with training set filtering, we do not use co-teaching and instead, we select MpMRI and M.S. MpMRI for experiments in this section. We use these two models to compute the uncertainty for all training data first, and then the filtering process can be done either patch-driven or patient-driven on the training set, as we discussed in Section \ref{evi_focal}.

In Table \ref{pat_res_patc_fil}, we present the \textbf{patient-based results} on filtering 10\% and 20\% of training \textit{patches} in the training set for the two selected models. Table \ref{pat_res_pati_fil} represents the same as Table \ref{pat_res_patc_fil}, except we filter 10\% and 20\% of training \textit{patients} in the training set. The results from both tables demonstrate that the MpMRI model performs better than the M.S.MpMRI model in each filtration rate, and the binary classification performance improves when filtering more uncertain data for both models. From the expected calibration error (ECE), we observed that the MpMRI model with 20\% filtering on both patch- and patient-based has a lower ECE value compared to those for the M.S. MpMRI model, demonstrating the predicted output probabilities of the MpMRI model matches well with the actual probabilities of the ground truth. Comparing these results with those from Table \ref{pat_res}, we can conclude that the dataset filtering method applied to the training set, together with the evidential focal loss we proposed, can effectively improve the classification performance. 

\begin{table}[!ht]
\centering
\begin{tabular}{c|c|c|c|c|c|c|c|cl}
                                       & \multicolumn{1}{c|}{Data}    & F.R. & F.M. & Acc.          & Sen.          & Spec.         & AUC           & ECE $\downarrow$ &  \\ \cline{1-9}

\multicolumn{1}{c|}{MpMRI}             & T2+ADC                      & 10\% & patch   & \textbf{82.4}          & \textbf{85.7}          & 81.5          & \textbf{85.7$\pm$9.9}          & 0.27 & \\
\multicolumn{1}{c|}{M.S.MpMRI}         & T2+ADC                      & 10\% & patch   & \textbf{82.4}          & 71.4          & \textbf{85.2}          & 83.6$\pm$13.5          & \textbf{0.15}  & \\ \hline
\multicolumn{1}{c|}{MpMRI}             & T2+ADC                      & 20\% & patch   & \textbf{85.3}          & \textbf{100}           & 81.5          & \textbf{98.4$\pm$1.6}          &  \textbf{0.20} & \\
\multicolumn{1}{c|}{M.S. MpMRI}        & T2+ADC                      & 20\% & patch   & \textbf{85.3}          & 71.4          & \textbf{88.9}          & 92.6$\pm$7.4          & 0.22 &  \\ \cline{1-9}
\end{tabular}
\caption{\textbf{Patient-based results} of experiments using evidential focal loss and \textit{patch-based filtering}. Standard deviations are computed from the 95\% bootstrap confidence interval with $n = 3000$ samples. The ``F.R.'' and ``F.M.'' represent the filtering rate and the filtering method, respectively. The best results for each filtration rate are \textbf{bold}. All units of the numeric values are in \%.}
\label{pat_res_patc_fil}
\end{table}

\begin{table}[!ht]
\centering
\begin{tabular}{c|c|c|c|c|c|c|c|cl}
                                       & \multicolumn{1}{c|}{Data}    & F.R. & F.M. & Acc.          & Sen.          & Spec.         & AUC           & ECE $\downarrow$ & \\ \cline{1-9}

\multicolumn{1}{c|}{MpMRI}             & T2+ADC                      & 10\% & patient & \textbf{88.2}          & \textbf{100}           & \textbf{85.2}          & \textbf{92.6$\pm$7.4}          & 0.27  & \\
\multicolumn{1}{c|}{M.S. MpMRI}        & T2+ADC                      & 10\% & patient & 85.3          & \textbf{100}           & 81.5          & 86.2$\pm$9.0          & \textbf{0.24} & \\ \hline
\multicolumn{1}{c|}{MpMRI}             & T2+ADC                      & 20\% & patient & \textbf{73.5}          & \textbf{85.7}          & 70.4          & \textbf{86.8$\pm$12.6}          & \textbf{0.21} & \\
\multicolumn{1}{c|}{M.S. MpMRI}        & T2+ADC                      & 20\% & patient & \textbf{73.5}          & 71.4          & \textbf{74.1}          & 84.6$\pm$14.2          & 0.22 &  \\ \cline{1-9}
\end{tabular}
\caption{\textbf{Patient-based results} of experiments using evidential focal loss and \textit{patient-based filtering}. Standard deviations are computed from the 95\% bootstrap confidence interval with $n = 3000$ samples. The ``F.R.'' and ``F.M.'' represent the filtering rate and the filtering method, respectively. The best results for each filtration rate are in \textbf{bold}. All units of the numeric values are in \%.}
\label{pat_res_pati_fil}
\end{table}

Moreover, an interesting observation reveals a gradual deterioration in performance with \textit{patient-driven} filtering. This phenomenon may stem from the distinction between patch-driven and patient-driven filtering approaches. In patch-driven filtering, the exclusion of training patches with high uncertainty values is straightforward during the training process, irrespective of the patient to which these patches belong. Conversely, in the case of patient-driven filtering, we have to consider the average uncertainty of the 20 patches for each patient. In cases where the average uncertainty for a patient falls below the defined threshold, all patches for that patient are used in training, irrespective of whether specific patches exhibit exceptionally high uncertainty. Consequently, there exists a risk of erroneously retaining patches with high uncertainty if the average uncertainty for the corresponding patient is relatively low, potentially impacting the model's performance. This nuanced difference contributes to the observed discrepancy in results, favoring the patch-driven filtration approach.

\textbf{Ablation Study: }As previously mentioned, Table \ref{pat_res} corresponds to experiments conducted without using evidential focal loss or filtering. On the other hand, Table \ref{pat_res_patc_fil} and \ref{pat_res_pati_fil} encompass experiments that incorporate both these elements. To solely examine the influence of our proposed loss, we conducted an experiment employing the evidential focal loss without any filtering (0\%). These results are summarized in Table \ref{abl_filter} in comparison with 20\% \textit{patch-based filtering} approach. As can be seen, even without any data filtering during the training, we could correctly classify all patients with clinically significant PCa (sensitivity = 100\%), which demonstrates a significant improvement compared to MpMRI using traditional focal loss in Table \ref{pat_res} and also the baseline result in \cite{grebenisan2021spatial}. As expected, the addition of data filtering further improves the results for all metrics. The original results based on image patches of Table \ref{abl_filter} can be found in Appendix \ref{app:pat_res}.

\begin{table}[h!]
\centering
\begin{tabular}{l|ccccc|ccccc}
                                & \multicolumn{5}{c|}{filter 0\%}                              & \multicolumn{5}{c}{filter 20\%}                              \\ \hline
                                & Acc. & Sen. & Spec. & AUC & ECE                                    & Acc. & Sen. & Spec. & AUC & ECE                                    \\ \hline
\multicolumn{1}{c|}{MpMRI}      & 82.4 & 100  & 77.8  & 89.4$\pm$9.1  & 0.29 & 85.3 & 100  & 81.5  & 98.4$\pm$1.6 & 0.20 \\ \hline
\multicolumn{1}{c|}{M.S. MpMRI} & 76.5 & 100 & 70.4  & 82.0$\pm$12.7                         & 0.23 & 85.3 & 71.4 & 88.9  & 92.6$\pm$7.4 & 0.22                          \\ \hline
\end{tabular}
\caption{Ablation on employing proposed evidential focal loss with and without data filtering for the two selected architectures. The \textbf{Patient-based} results are reported. Standard deviations are computed from the 95\% bootstrap confidence interval with $n = 3000$ samples. The best results for each filtration rate are in \textbf{bold}. All units of the numeric values are in \%.}
\label{abl_filter}
\end{table}

\subsection{Filtering during deployment}

So far, our exploration has centered on the impact of the data filtering strategy during training, aimed at enhancing model robustness and performance. It is also possible to apply filtering on the test set, i.e., when deploying the model to real clinical routines. This is equivalent to refraining from making decisions on the test samples that are identified as highly uncertain. The evaluation involves utilizing the pre-trained MpMRI and M.S. MpMRI models, each with 0\% and 20\% \textit{patch-based} training filtering rates, as final models, and assessing their performance on the test set. The performance on filtering the test data based on different uncertainty thresholds using pre-trained models with 20\% filtering during training is shown in Figure \ref{test_filter_res}. For the performance of the other two models (0\% filtering), refer to Appendix \ref{filter_deploy_appd}. We started with an uncertainty threshold of 1.0 (keeping patients with \textit{ncertainty < 1.0}); we progressively adjusted the threshold to 0.68, 0.63, 0.58, and ended at a threshold value of 0.53. These thresholds are determined based on the uncertainty value computed in the test set. We observe that the model improves its performance when filtering out highly uncertain patients from the test set, and eventually classified all patients correctly, as shown in Figure \ref{fig:test_mpmri20}. 

This pragmatic approach bears relevance in real clinical settings, offering radiologists an efficient means to allocate their time, focusing on patients filtered out during the diagnostic process due to their high uncertainty values, rather than those confidently classified.
% In the clinical practice, our model could help radiologists to save lots of time on well-classified patients and focus more on the patients that have been filtered out (with high uncertainty value) during the diagnostic process.

\begin{figure}[h!]
     \centering
     \begin{subfigure}[h]{0.49\textwidth}
         \centering
         \includegraphics[width=\textwidth]{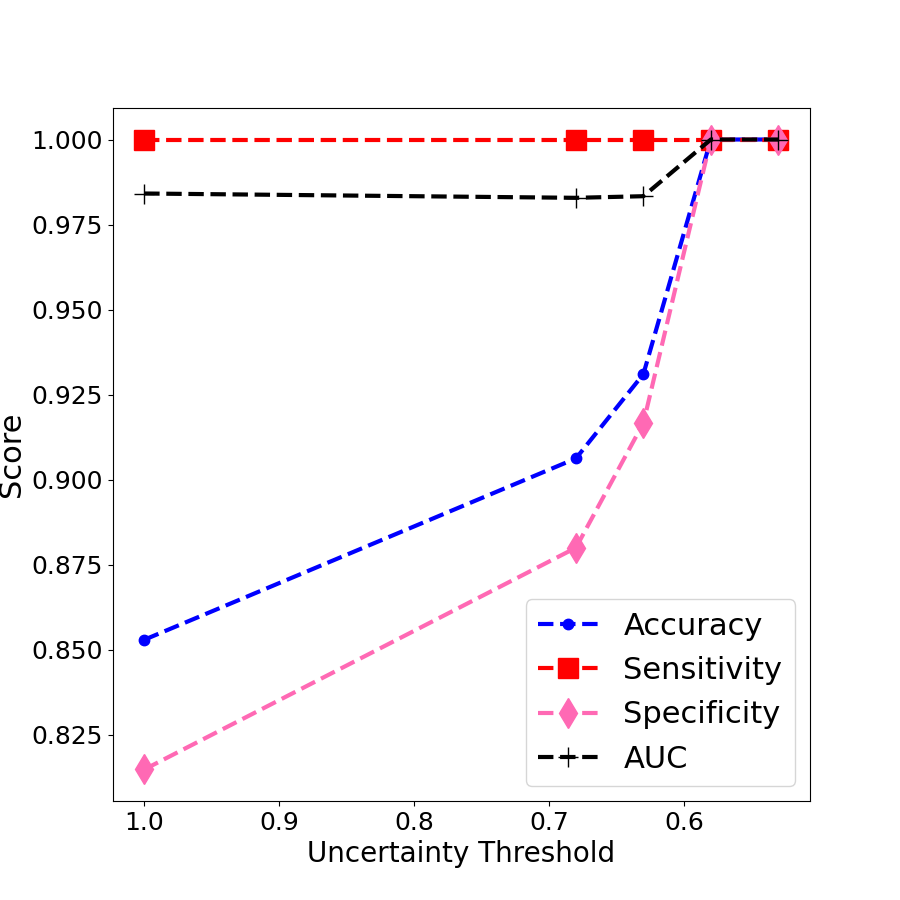}
         \caption{MpMRI, with 20\% filtering on training patches}
         \label{fig:test_mpmri20}
     \end{subfigure}
     \hfill
     \begin{subfigure}[h]{0.49\textwidth}
         \centering
         \includegraphics[width=\textwidth]{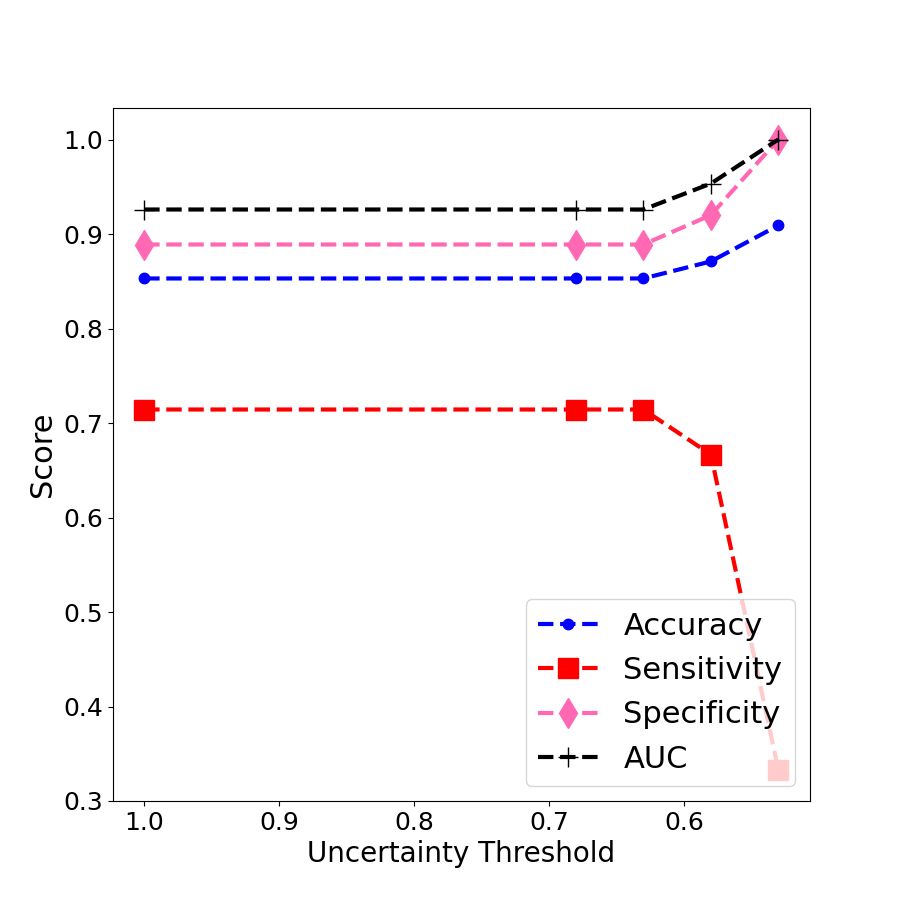}
         \caption{M.S. MpMRI, with 20\% filtering on training patches}
         \label{fig:test_multi20}
     \end{subfigure}
     \caption{Test performance of selected models based on uncertainty threshold.}
     \label{test_filter_res}
\end{figure}

\section{Conclusion}

In this study, we introduced a novel approach for unpaired image-to-image translation of prostate mp-MRI and developed a robust deep-learning model for classifying clinically significant PCa using evidential focal loss. We demonstrated the effectiveness of our method on our local dataset, reinforced by a publicly available one, and showed that uncertainty-aware filtering during both training and deployment can significantly improve the PCa classification performance. The quantitative results for image translation also demonstrated that our proposed domain transfer framework using the ACL-GAN model outperforms the baseline SDNet \cite{grebenisan2021spatial}. Our approach holds promise in expediting the diagnostic process by identifying patients with high uncertainty, allowing clinicians to focus on precise diagnosis and expedite cases with high prediction certainty.

While our approach has shown promising results, there are still opportunities for improvement. One potential area for future work is to consider the spatial dependency between slices in volumetric MRI. Currently, our domain transfer framework only accepts 2D images as input and output, and we reshape the volumetric MRI into several 2D slices. However, explicitly splitting 3D images into 2D slices may eliminate the spatial dependency within each MRI data and affect the classification results. Therefore, a plausible solution could be translating the entire 3D MRI volume from 3.0T to 1.5T instead of handling individual slices.

Lastly, there is great potential for further improving the classification performance by combining more images from different MRI functional sequences, such as b-value and K$^{trans}$. We have already demonstrated that incorporating additional ADC images significantly enhances classification performance. We believe that if we successfully translate other images from b-value or K$^{trans}$ acquired at 3.0T to 1.5T and incorporate them into the classification, the results could be further improved. However, the additional MRI sequences may not be available in the local 1.5T dataset. The conversion process may become feasible if we acquire those sequences from local hospitals.

% \disclosures 
\subsection*{Disclosures}
We declare we don't have conflicts of interest.

\subsection* {Code, Data, and Materials Availability} 
ProstateX data is publicly available here \footnote{\url{https://wiki.cancerimagingarchive.net/pages/viewpage.action?pageId=23691656}}, our local KHSC data is not available due to ethical concerns. Our code is at \url{https://github.com/med-i-lab/DT_UE_PCa}, we will make the code available upon acceptance.

\subsection* {Acknowledgments}
This work was supported by the Natural Sciences and Engineering Council of Canada, Canadian Institutes for Health Research, and Queen's University. Parvin Mousavi is supported by Canada CIFAR AI Chair and the Vector AI Institute.

%%%%% References %%%%%

\bibliography{report}   % bibliography data in report.bib
\bibliographystyle{spiejour}   % makes bibtex use spiejour.bst

%%%%% Appendix %%%%%%%%%

\newpage

% Appendix is optional
\appendix

\section{Experimental Details} \label{app:exp_det}

In this section, we provide details of hyperparameter settings to train uncertainty-aware classification models described in Section \ref{setup}.

To train the ``MpMRI'' model for patch-driven filtering, we set the learning rate to 0.0001; weight decay to 0.01; total training epochs to 300, and batch size to 10. The class weights $\boldsymbol{\beta}$ in Equation \eqref{final_efl} are set to [0.25, 0.75] for filtering 10\%, and [0.25, 1.25] for filtering 20\% of the training data. For patient-driven filtering, all parameters are the same except the class weights $\boldsymbol{\beta}$ are set to [0.25, 1] for filtering 10\% of the training data. Last but not least, we set the initial learning rate to 0.0001; total training epochs to 300; batch size to 10; the learning rate decayed by a factor of 0.1 for every 200 epochs, and the class weights $\boldsymbol{\beta}$ are set to [0.25, 1] for filtering 20\% of the training data.

To train the ``M.S. MpMRI'' model for patch-driven filtering, we set the initial learning rate to 0.0001; weight decay to 0.01; total training epochs to 300. The class weights $\boldsymbol{\beta}$ in Equation \eqref{final_efl} are set to [0.25, 1] for both filtering 10\% and 20\% of the training data. For patient-driven filtering, all parameters were the same except the class weights $\boldsymbol{\beta}$ are set to [0.25, 1] for filtering 10\% and [0.25, 1.25] for filtering 20\% of the training data.

\section{Original Results} \label{app:pat_res}

In this section, we provide all original results based on image patches (Patch-based results) of all experiments we performed in the paper. We start by presenting Table \ref{app:pact_res}, which is the patch-based result for Table \ref{pat_res} in Section \ref{pca_res_nofil}.

\begin{table}[!ht]
\centering
\begin{tabular}{c|c|c|c|c|cl}
                                       & \multicolumn{1}{c|}{Data}    & Acc.          & Sen.          & Spec.         & AUC           &  \\ \cline{1-6}
\multicolumn{1}{c|}{SDNet \cite{grebenisan2021spatial}(baseline)}   & T2 & 77.8          & 27.9          & \textbf{90.7} & 74.8$\pm$4.1          &  \\
\multicolumn{1}{c|}{T2-only}           & \multicolumn{1}{c|}{T2}    & 65.6          & \textbf{75.7} & 63.0          & 77.9$\pm$3.9          &  \\ \hline
\multicolumn{1}{c|}{MpMRI}             & \multicolumn{1}{c|}{T2+ADC}    & 75.1          & 69.3          & 76.7          & 79.9$\pm$3.8          &  \\
\multicolumn{1}{c|}{Vol. MpMRI}        & \multicolumn{1}{c|}{T2+ADC}    & 64.9          & 62.9          & 65.4          & 63.4$\pm$5.5          &  \\
\multicolumn{1}{c|}{M.S. MpMRI}        & \multicolumn{1}{c|}{T2+ADC}    & 70.4          & 74.3          & 69.4          & 80.1$\pm$3.6          &  \\
\multicolumn{1}{c|}{MpMRI+co-teaching} & \multicolumn{1}{c|}{T2+ADC}    & \textbf{78.4} & 72.1          & 80.0          & \textbf{80.7$\pm$3.9} &  \\ \cline{1-6}
\end{tabular}
\caption{\textbf{Patch-based results} of experiments using conventional training paradigm in Section \ref{setup}. Standard deviations are computed from the 95\% bootstrap confidence interval with $n = 3000$ samples. The best results are \textbf{bold}. All units of the numeric values are in \%.}
\label{app:pact_res}
\end{table}

Then, we provide the patch-based results for patch-based filtering in Table \ref{app:patc_res_fil}, which corresponds to Table \ref{pat_res_patc_fil} in Section \ref{pca_res_fil}.

\begin{table}[!ht]
\centering
\begin{tabular}{c|c|c|c|c|c|c|c|cl}
                                       & \multicolumn{1}{c|}{Data}    & F.R. & F.M. & Acc.          & Sen.          & Spec.         & AUC           & ECE $\downarrow$ & \\ \cline{1-9}

\multicolumn{1}{c|}{MpMRI}             & T2+ADC                      & 10\% & patch & 77.5          & \textbf{80.7} & 76.7          & \textbf{82.4$\pm$3.4}          & 0.22 &  \\
\multicolumn{1}{c|}{M.S.MpMRI}         & T2+ADC                      & 10\% & patch & \textbf{79.7}          & 70.7          & \textbf{82.0}          & 81.1$\pm$3.8          & \textbf{0.16} &  \\ \hline
\multicolumn{1}{c|}{MpMRI}             & T2+ADC            & 20\% & patch & \textbf{83.8} & \textbf{80.0}  & \textbf{84.8} & \textbf{89.7$\pm$2.6}         & \textbf{0.16} &  \\
\multicolumn{1}{c|}{M.S. MpMRI}        & T2+ADC                      & 20\% & patch  & 82.4          & 73.6          & 84.6          & 87.6$\pm$2.8         & 0.17 & \\ \cline{1-9}
\end{tabular}
\caption{\textbf{Patch-based results} of experiments using evidential focal loss and \textit{patch-based filtering}. Standard deviations are computed from the 95\% bootstrap confidence interval with $n = 3000$ samples. The ``F.R.'' and ``F.M.'' represent the filtering rate and the filtering method, respectively. The best results for each filtration rate are \textbf{bold}. All units of the numeric values are in \%.}
\label{app:patc_res_fil}
\end{table}

For patient-based filtering, we provide the patch-based results in Table \ref{app:pat_res_pati_fil}, which corresponds to Table \ref{pat_res_pati_fil} in Section \ref{pca_res_fil}.

\begin{table}[!ht]
\centering
\begin{tabular}{c|c|c|c|c|c|c|c|cl}
                                       & \multicolumn{1}{c|}{Data}    & F.R. & F.M. & Acc.          & Sen.          & Spec.         & AUC           & ECE $\downarrow$ &  \\ \cline{1-9}

\multicolumn{1}{c|}{MpMRI}             & T2+ADC                      & 10\% & patient & 75.3          & \textbf{80.7} & 73.9   & \textbf{86.1$\pm$2.9}   & 0.21 &  \\
\multicolumn{1}{c|}{M.S. MpMRI}        & T2+ADC                      & 10\% & patient & \textbf{75.6}          & 75.6          & \textbf{76.9} & 82.4$\pm$3.6     & \textbf{0.17} & \\ \hline
\multicolumn{1}{c|}{MpMRI}             & T2+ADC  & 20\% & patient & 72.5  & \textbf{74.3} & 72.0          & \textbf{81.5$\pm$3.6}   & 0.22 &  \\
\multicolumn{1}{c|}{M.S. MpMRI}        & T2+ADC    & 20\% & patient & \textbf{76.3} & \textbf{74.3}     & \textbf{76.9} & 80.1$\pm$4.2    & \textbf{0.18} &  \\ \cline{1-9}
\end{tabular}
\caption{\textbf{Patch-based results} of experiments using evidential focal loss and \textit{patient-based filtering}. Standard deviations are computed from the 95\% bootstrap confidence interval with $n = 3000$ samples. The ``F.R.'' and ``F.M.'' represent the filtering rate and the filtering method, respectively. The best results for each filtration rate are in \textbf{bold}. All units of the numeric values are in \%.}
\label{app:pat_res_pati_fil}
\end{table}

\noindent Table \ref{abl_fil_pat} shows the original patch-based results based on image patches for the ablation study conducted in Section \ref{pca_res_fil}.

\begin{table}[h!]
\centering
\begin{tabular}{l|ccccc|ccccc}
           & \multicolumn{5}{c|}{filter 0\%}                                  & \multicolumn{4}{c}{filter 20\%}                                  \\ \hline
           & Acc. & Sen. & Spec. & AUC & ECE                                        & Acc. & Sen. & Spec. & AUC & ECE                                        \\ \hline
MpMRI      & 74.1 & 79.3 & 72.8  & 84.8$\pm$2.9 & 0.21 & 83.8 & 80.0 & 84.8  & 89.7$\pm$2.6 & 0.16 \\ \hline
M.S. MpMRI & 73.7 & 86.4 & 70.4  & 80.4$\pm$3.1 & 0.22                              & 82.4 & 73.6 & 84.6  & 87.6$\pm$2.8 & 0.17                               \\ \hline
\end{tabular}
\caption{Patch-based results for disable filtration on the training set for two models.}
\label{abl_fil_pat}
\end{table}

we also provide the visualization of patch-based AUC curves for the selected experiments in Section \ref{pca_res_nofil} and \ref{pca_res_fil}, along with the 95\% confidence interval against the baseline model in Figure \ref{pac_vis}. 

% We conclude that our best model ``MpMRI-EFL'' with 20\% filtration on the training set has a tighter confidence bound compared to other models, indicating the best model is more precise on the test set against all other methods, particularly to the baseline method.

\begin{figure}[h!]
     \centering
     \begin{subfigure}[h]{0.49\textwidth}
         \centering
         \includegraphics[width=\textwidth]{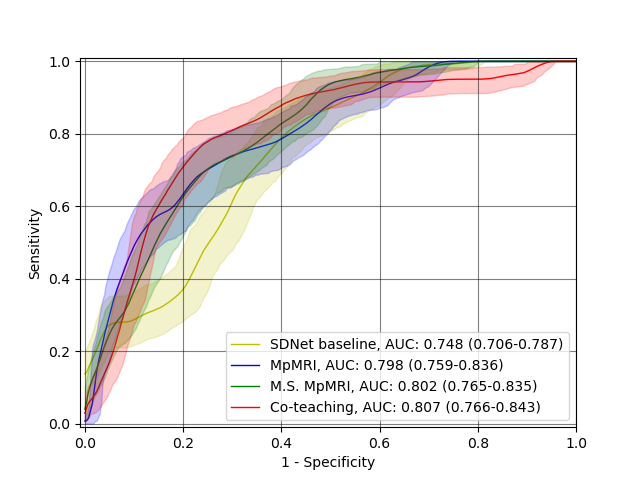}
         %\caption{Comparison between the baseline and the models without filtration}
         \caption{}
         \label{fig:auc_old}
     \end{subfigure}
     \hfill
     \begin{subfigure}[h]{0.49\textwidth}
         \centering
         \includegraphics[width=\textwidth]{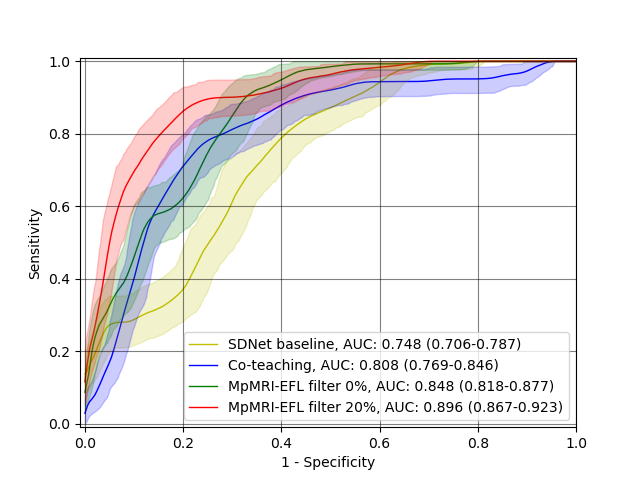}
         %\caption{Comparison between the baseline, the best model without filtration, and the best model with filtration on the training set.}
         \caption{}
         \label{fig:auc_new}
     \end{subfigure}
     \caption{Both figures demonstrate the original AUC results. \ref{fig:auc_old} shows the comparison of AUC curves between the baseline and the models without filtration (experiments in the first category); \ref{fig:auc_new} shows the comparison of AUC curves between the baseline, the best model without filtration, and the best model with filtration on the training set. ``EFL'' is short for Evidential Focal Loss. The shaded areas in both figures represent the 95\% confidence intervals (CI) of each model. CIs are obtained by using Bootstrap with $n = 3000$.}
     \label{pac_vis}
\end{figure}

\section{Filtration while deploying} \label{filter_deploy_appd}

Next, we provide the performance for test set filtering using pre-trained MpMRI and M.S. MpMRI with 0\% filtering rate on the training set in Figure \ref{test_filter_res_appd}.  
\begin{figure}[htbp]
     \centering
     \begin{subfigure}[h]{0.49\textwidth}
         \centering
         \includegraphics[width=\textwidth]{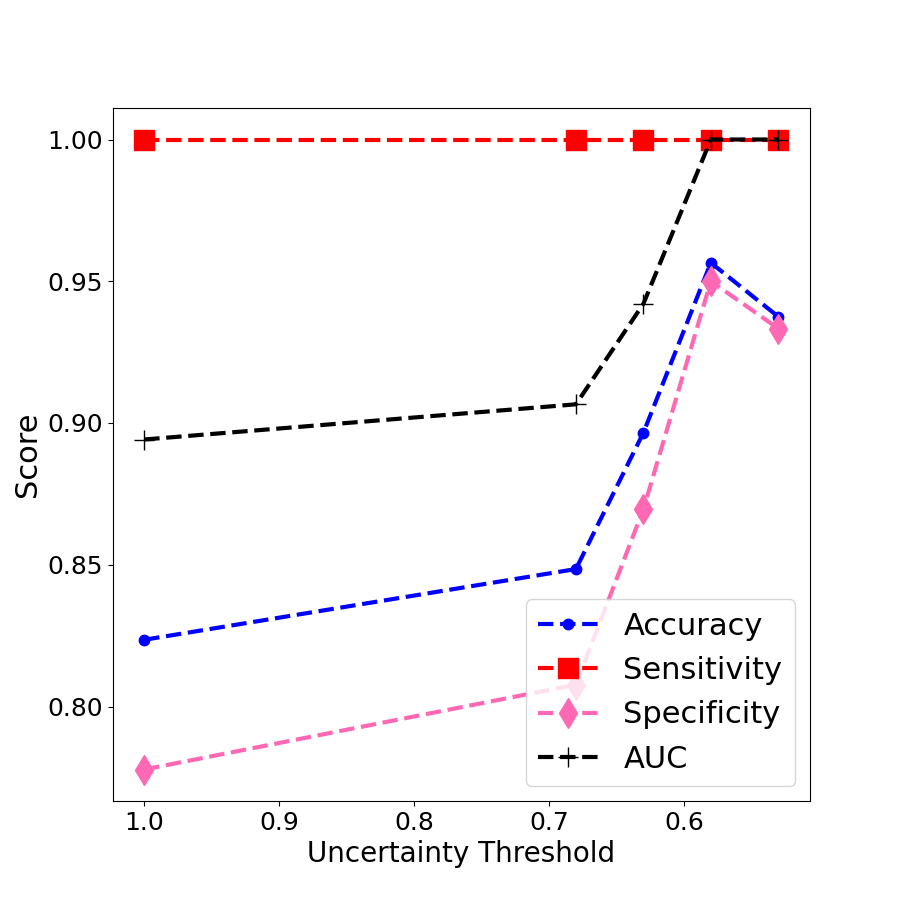}
         \caption{MpMRI, with 0\% filtering on training patches}
         \label{fig:test_mpmri0}
     \end{subfigure}
     \hfill
    %  \begin{subfigure}[h]{0.49\textwidth}
    %      \centering
    %      \includegraphics[width=\textwidth]{Figures/test_filter_analysis_PL_mpmri20_new_temp.png}
    %      \caption{}
    %      \label{fig:test_mpmri20}
    %  \end{subfigure}
    %  \hfill
     \begin{subfigure}[h]{0.49\textwidth}
         \centering
         \includegraphics[width=\textwidth]{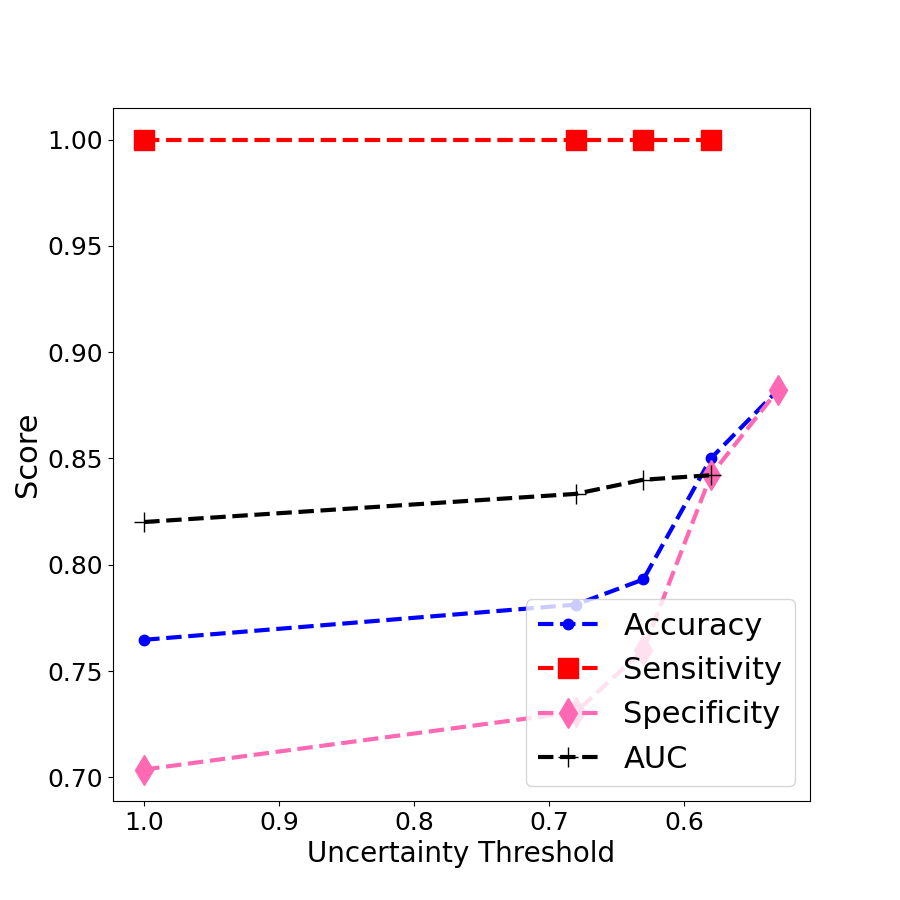}
         \caption{M.S. MpMRI, with 0\% filtering on training patches}
         \label{fig:test_multi0}
     \end{subfigure}
    %  \hfill
    %  \begin{subfigure}[h]{0.49\textwidth}
    %      \centering
    %      \includegraphics[width=\textwidth]{Figures/test_filter_analysis_PL_multihead20_new_temp.png}
    %      \caption{}
    %      \label{fig:test_multi20}
    %  \end{subfigure}
     \caption{Test performance of selected models based on uncertainty threshold.}%Test set performance of filtering 0\% to 40\% of test data on the selected models. \ref{fig:test_mpmri0} represents the test performance of ``MpMRI'' model with 0\% filtration on the training set, and \ref{fig:test_multi0} represents the test performance of ``M.S. MpMRI'' model with 0\% filtration on the training set.}
     \label{test_filter_res_appd}
\end{figure}

%%%%% Biographies of authors %%%%%

\newpage
\vspace{2ex}\noindent\textbf{First Author} insert biography

\vspace{1ex}
\noindent Biographies and photographs of the other authors are not available.

\listoffigures
\listoftables

\end{spacing}
\end{document}